\documentclass[%
 reprint,
 amsmath,amssymb,
 aps,
]{revtex4-2}

\usepackage{breqn}
\usepackage{amsmath}
\usepackage{color}
\usepackage{graphicx}
\usepackage{dcolumn}
\usepackage{bm}
\usepackage{hyperref}

\hypersetup{
    colorlinks=true,
    linkcolor=blue,
    citecolor=blue,
    filecolor=magenta,      
    urlcolor=blue,
    pdftitle={Overleaf Example},
    pdfpagemode=FullScreen,
    }

\begin{document}

\preprint{APS/123-QED}

\title{Exploring the Dynamics of General Relativistic Binary-Single and Binary-Binary Encounters of Black Holes}

\author{Felix M. Heinze$^{1}$}
\author{Bernd Brügmann$^{1}$}

\author{Tim Dietrich$^{2,3}$}
\author{Ivan Markin$^{3}$}
\affiliation{$^{1}$Friedrich-Schiller-Universität Jena$,$ Theoretisch Physikalisches Institut$,$ 07743 Jena$,$ Germany}
\affiliation{$^{2}$Max-Planck-Institut für Gravitationsphysik (Albert-Einstein-Institut)$,$ 14476 Potsdam$,$ Germany}
\affiliation{$^{3}$Universität Potsdam$,$ 14469 Potsdam$,$ Germany}
\date{\today}

\begin{abstract}
In this exploratory study, we demonstrate the capability of the numerical-relativity code BAM to simulate fully relativistic black-hole binary-single and binary-binary encounters. While previous work has demonstrated the general capability of numerical-relativity frameworks to evolve spacetimes with $N$ black holes, detailed explorations of such encounters remain limited. We focus on scenarios involving initially non-spinning, equal-mass black holes that result in a variety of dynamical outcomes, including flybys, delayed or accelerated eccentric mergers, exchanges, and more complex interactions. Our results show that we can reliably simulate scattering experiments involving three and four black holes, which exhibit interesting dynamics and gravitational-wave signals. The dynamics of these systems show noticeable differences compared to analogous systems in post-Newtonian approximations up to 2.5PN. A key result is that the gravitational waveforms exhibit remarkable features that could potentially make them distinguishable from regular binary mergers.
\end{abstract}

                              
\maketitle


\section{\label{sec:introduction}Introduction}

The motion of $N$ bodies under their mutual gravitational interaction has been a central problem in astrophysics and celestial mechanics for centuries, dating back to 1687 when Isaac Newton published his \textit{Principia} \cite{Newton1687}. While two-body systems are well understood in the framework of Newtonian point particle mechanics, the inclusion of more than two bodies introduces a level of complexity that defies analytic solutions. Such $N$-body systems exhibit highly nonlinear and chaotic behavior, where small variations in the initial conditions can lead to vastly different outcomes, making long-term predictions very difficult. For the problem of $N$ black holes moving in close proximity and at high speeds, the Newtonian and post-Newtonian approximations can become highly inaccurate, and it is therefore necessary to solve the complete field equations of general relativity, which introduces additional complexity to the dynamics.

Interactions of $N$ black holes are expected to be common in environments with high stellar densities, such as the dense cores of globular clusters or nuclear clusters, where mass segregation leads to a high central density of massive objects that can even lead to observable direct collisions of stars \cite{Lombardi1996, Fregeau2002}. This makes it necessary to consider black hole binaries having close encounters with single black holes or with other binaries \cite{Gultekin2003, Miller2002}. Furthermore, successive galaxy mergers can even lead to close encounters of multiple supermassive black holes \cite{Valtonen1994, Valtonen1996, Hoffman2007, Pau2010}. Such systems have already been observed, for example in the triple quasar QQQ 1429-008 \cite{Djorgovski2007}.

Binary-single and binary-binary scattering processes have been studied extensively in the past. In the first studies, a large number (up to more than a million) of numerical simulations of such scattering events have been performed in the Newtonian point particle limit, being mostly concerned about closest-approach distances, the distribution of outcome properties, the cross-sections for ionization, exchange, and resonance scattering, as well as the energy exchange as a function of the orbital parameters and the relative incoming velocity \cite{Hills1975, Hut1983, Hoffer1983, Mikkola1983a}. Another key result of the study of these interactions is Heggie's law \cite{Heggie1975} which states that binaries with a binding energy that exceeds the average kinetic energy of the other cluster members (called \textit{hard} binaries) tend to become harder through such scattering processes, whereas \textit{soft} binaries (where the binding energy of the binary is smaller than the average kinetic energy of the other cluster members) tend to become softer over time. The exchange of energy between a hard binary and other cluster members is able to heat the cluster, providing an important mechanism to prevent core collapse or accelerate evaporation \cite{Heggie1979, Dokuchaev1981, Mikkola1983b}. Later studies of binary-single and binary-binary scatterings accounted for additional effects, such as gravitational radiation, tidal interactions, and physical collisions \cite{McMillan1986, Cleary1990, Fregeau2004}. More recently, post-Newtonian terms up to the 2.5th order have been incorporated \cite{Samsing2018a, Samsing2018b, Codazzo2024}. It has been demonstrated multiple times that three-body and four-body interactions are efficient at producing binary black hole mergers \cite{Morscher2015, Rodriguez2016, Samsing2017a, Trani2019} and that they can also be a possible source of eccentric mergers \cite{Samsing2014, Samsing2017b, Samsing2018c, Zevin2019, ArcaSedda2021, DallAmico2024}. Interactions between multiple supermassive black holes can significantly reduce their merger time, providing a potential solution to the final parsec problem \cite{Ryu2018}.

In full general relativity, prior to the $N$-body problem, even the 2-body problem required substantial developments. Early examples for non-axisymmetric black-hole binaries using the ``puncture method'' include grazing collisions \cite{Alcubierre2000} and the first complete binary orbit before merger \cite{Brugmann2004}. The year 2005 marks a breakthrough for the late inspiral and merger of two black holes \cite{Pretorius2005, Campanelli2005, Baker2005}, where either dynamical black-hole excision or a ``moving puncture'' approach was used. Such methods generalize directly to $N$ black holes, as first demonstrated for moving punctures in \cite{Lousto2008}. After that, several fully relativistic studies of systems containing more than two black holes have been carried out, but actually not that many. These studies have either looked at arbitrary $N$-black-hole configurations \cite{Lousto2008, Galaviz2010} (recently \cite{Bamber2025} evolved a cluster of 25 black holes), or they focused on the dynamics of systems containing a black hole orbiting in close proximity to a black-hole binary \cite{Campanelli2008, Ficarra2023, Ficarra2024}, but more general binary-single or binary-binary encounters have so far not been investigated in detail.

Since small changes in the initial conditions can lead to vastly different outcomes, it often does not suffice to look at only a few similar scattering configurations. Most of the major results for $N$-body scattering processes are obtained by running a large number of simulations (often more than hundreds of thousands or millions) and investigating the statistical properties of the outcomes. This poses a major challenge for numerical-relativity simulations as they are computationally expensive, such that it is unfeasible to produce even on the order of a thousand of such scattering experiments within a reasonable time.

In this paper, we focus on performing a few selected case studies in which we qualitatively analyze the results, demonstrating the possibility of performing accurate relativistic black-hole binary-single and binary-binary scattering simulations using the numerical-relativity code \textsc{BAM}. We present for the first time a set of simulations of such encounters with more realistic initial conditions (i.e.\ large initial separations of the scattering objects), including one black-hole binary-binary and five binary-single scattering experiments. As a first step toward exploring the vast parameter space, we focus on the simple scenario of initially non-spinning, equal-mass black holes and (quasi-)circular binaries encountering another single black hole or binary in two possible orientations.

The paper is structured as follows. In Section \ref{sec:methods} we provide a quick summary of how we compute the initial data and how these are evolved in time, including the numerical methods we use to solve the corresponding equations. In Section \ref{sec:exp_setup} we explain how we set up our binary-single and binary-binary scattering experiments and what needs to be taken into consideration in order to obtain accurate results. Our first simulations are presented in Section \ref{sec:results}, where we show examples of different outcomes of the encounters, their dynamics, as well as their gravitational-wave signals. A summary of our results can be found in Section \ref{sec:summary}. In this paper we use metric signature $(-,+,+,+)$, geometric units where $G=c=1$, and a mass scale $M$ which in code units is $M=1$.

\section{\label{sec:methods}Methods}
Numerical relativity offers a variety of methods for the solution of the Einstein equations (e.g.\ \cite{bona2009elements, Alcubierre2008, Baumgarte2010, Gourgoulhon2012, Bruegmann2018}). Here, we adopt a specific approach suitable for vacuum spacetimes with N black holes, based on the standard 3+1 decomposition. For $N>2$, the puncture method provides a flexible framework: black hole punctures can move freely across a numerical grid without requiring adapted coordinates or excision. This section introduces the relevant aspects of the moving puncture method for this work.

A 3+1 decomposition of the Einstein equations of general relativity yields a set of constraint and evolution equations. The constraint equations can be solved to obtain initial data that describe the geometry of a valid 3-dimensional space-like hypersurface of a general (3+1)-dimensional spacetime, and the evolution equations govern the time evolution of the associated geometric quantities. The primary geometric quantities that describe the geometry of such a spatial slice are the induced 3-metric $\gamma_{ij}$, from which the intrinsic Riemann curvature can be computed, and the extrinsic curvature $K_{ij}$, which describes how the 3D hypersurface is embedded in the surrounding 4D spacetime.

\subsection{Initial Data}
In vacuum, the constraint equations are
\begin{equation}
    D_j (K^{ij} - \gamma^{ij} K) = 0,
    \label{eq:mom_constraint}
\end{equation}
\begin{equation}
    R^2 + K^2 - K_{ij} K^{ij} = 0,
    \label{eq:ham_constraint}
\end{equation}
where $K$ is the trace of the extrinsic curvature, $R$ is the 3D Ricci scalar, and $D_j$ is the 3D covariant derivative associated with $\gamma_{ij}$. These constraints can be solved by using the puncture method for black holes \cite{Brandt1997}. Initial data for $N$ black holes are modeled by adopting the Brill-Lindquist wormhole topology with $N + 1$ asymptotically flat ends and $N$ connecting "throats". The ends are compactified and identified with points on $\mathbb{R}^3$. The coordinate singularities at the points resulting from compactification are called \textit{punctures}.

We use a conformal transverse-traceless decomposition of the metric and the extrinsic curvature, assuming
\begin{align}
    \gamma_{ij} &= \psi^4 \tilde{\gamma}_{ij}, \\
    K_{ij} &= \psi^{-2} \bar{A}_{ij} + \frac{1}{3} K \gamma_{ij}, 
\end{align}
where $\bar{A}_{ij}$ is trace-free. We further adopt Cartesian coordinates, assume conformal flatness ($\tilde{\gamma}_{ij} = \eta_{ij}$) and maximal slicing ($K=0$), which greatly simplifies and decouples the constraint equations for $\psi$ and $\bar{A}_{ij}$. With these choices, the momentum constraint (\ref{eq:mom_constraint}) takes the form
\begin{align}
    \partial_j \bar{A}^{ij} = 0.
\end{align}
Bowen and York \cite{BowenYork1980} found a non-trivial analytic solution to the momentum constraint, which describes $N$ black holes, where the $\alpha$-th black hole has an ADM linear momentum $P_\alpha^i$ and angular momentum $J_\alpha^i$, given by
\begin{dmath}
    \bar{A}^{ij} = \sum_{\alpha=1}^N \left[\frac{3}{2r_{C_\alpha}^2} \left( P_\alpha^i l_{C_\alpha}^j + P_\alpha^j l_{C_\alpha}^i - (\eta^{ij} - l_{C_\alpha}^i l_{C_\alpha}^j) l_{C_\alpha}^k P_{\alpha,k} \right) + \frac{6}{r_{C_\alpha}^3} l_{C_\alpha}^{(i} \epsilon^{j)kl} J_{\alpha,k} l_{C_\alpha, l} \right].
\end{dmath}
Here $r_{C_\alpha}  = |x^i - C_\alpha^i|$ and $l_{C_\alpha}^i = (x^i - C_\alpha^i)/r_{C_\alpha}$, where $C_\alpha^i$ are the coordinates of the center of the $\alpha$-th black hole. \\
The initial conformal factor $\psi$ can be obtained by solving the Hamiltonian constraint (\ref{eq:ham_constraint}) which now takes the form of the nonlinear elliptic equation
\begin{equation}
    \Delta \psi + \frac{1}{8} \psi^{-7} \bar{A}_{ij} \bar{A}^{ij} = 0.
    \label{eq:ham_constraint_psi}
\end{equation}
This can be solved numerically. $\psi$ is usually split into a sum of a singular term (containing the time-symmetric Brill-Lindquist solution of $N$ black holes without spin and linear momentum \cite{BrillLindquist1963}) and a finite correction $u$,
\begin{equation}
    \psi = 1 + \sum_{\alpha = 1}^N \frac{m_\alpha}{2r_\alpha} + u = \psi_{BL} + u,
    \label{eq:psi_split}
\end{equation}
with $u \rightarrow 0$ 
as $r \rightarrow \infty$. Here, the parameter $m_\alpha$ is the \textit{bare mass} of the $\alpha$-th puncture, which is only equal to the ADM mass or irreducible mass of the $\alpha$-th black hole in the limit of infinite separation.

\subsection{Time Evolution}
In order to evolve the initial data in time, we follow the ``moving puncture'' method as described in \cite{Brugmann2008}. For recent discussions of puncture evolutions, see e.g.\ \cite{Daszuta2021, Etienne2024, Ficarra2025}. We use the $\chi$-method and solve the BSSN evolution system which evolves the BSSN variables $\tilde{\gamma}_{ij}$, K,
\begin{align}
    \chi &= \psi^{-4}, \\
    \tilde{A}_{ij} &= \psi^{-6} \bar{A}_{ij}, \\
    \tilde{\Gamma}^i &= \tilde{\gamma}^{jk} \tilde{\Gamma}^i_{jk} = -\partial_j \tilde{\gamma}^{ij}
\end{align}
according to the evolution equations
\begin{align}
\begin{split}\label{eq:1}
    (\partial_t - \mathcal{L}_\beta)\chi ={}& \frac{2}{3} \chi (\alpha K - \partial_j \beta^j),
\end{split}\\
    (\partial_t - \mathcal{L}_\beta) \tilde{\gamma}_{ij} ={}& -2\alpha \tilde{A}_{ij}, \\
    (\partial_t - \mathcal{L}_\beta) K ={}& -\tilde{D}^i \tilde{D}_i \alpha + \alpha (\tilde{A}_{ij} \tilde{A}^{ij} + \frac{1}{3} K^2), \\
\begin{split}\label{eq:2}
    (\partial_t - \mathcal{L}_\beta) \tilde{A}_{ij} ={}& \chi [-\tilde{D}_i \tilde{D}_j \alpha + \alpha R_{ij}]^{TF} \\
    &+ \alpha \left[K \tilde{A}_{ij} - 2 \tilde{A}_{ik} \tilde{A}^k{}_j \right],
\end{split}\\
\begin{split}\label{eq:3}
    \partial_t \tilde{\Gamma}^i ={}& \tilde{\gamma}^{jk} \partial_j \partial_k \beta^i + \frac{1}{3} \tilde{\gamma}^{ij} \partial_j \partial_k \beta^k + \beta^j \partial_j \tilde{\Gamma}^i \\
    &- \tilde{\Gamma}^j \partial_j \beta^i + \frac{2}{3} \tilde{\Gamma}^i \partial_j \beta^j -2\tilde{A}^{ij} \partial_j \alpha \\
    &+ 2\alpha (\tilde{\Gamma}^i_{jk} \tilde{A}^{jk} - \frac{2}{3} \tilde{A}^{ij} \partial_j \ln \chi \\
    &- \frac{2}{3} \tilde{\gamma}^{ij} \partial_j K),
\end{split}
\end{align}
where $\tilde{D}_i$ is the 3D covariant derivative with respect to $\tilde{\gamma}_{ij}$ and $TF$ denotes the trace-free part of the expression with respect to $\gamma_{ij}$.

\subsection{Gauge Choices}
The lapse $\alpha$ and the shift $\beta^i$, which encode the gauge freedom in choosing space and time coordinates, are chosen such that the initial values are
\begin{align}
    \alpha &= \psi_{BL}^{-2}, \\ \beta^i &= 0,
\end{align}
and they are evolved according to the evolution equations
\begin{align}
    (\partial_t - \beta^j \partial_j) \alpha ={}& -2 \alpha K, \\
    (\partial_t - \beta^j \partial_j) \beta^i ={}& \frac{3}{4} B^i, \\
    (\partial_t - \beta^j \partial_j) B^i ={}& (\partial_t - \beta^j \partial_j) \tilde{\Gamma}^i - \eta B^i,
\end{align}
where $\eta$ is a constant with a typical value of $1/(2M)$ with $M$ being the total mass of the spacetime. These equations describe the 1+log slicing and the gamma-driver conditions with shift advection terms added to all the time derivatives. It has been demonstrated that these gauge conditions lead to a well-posed initial-value problem \cite{Gundlach2006}, which allows for long-term stable numerical evolutions of $N$ black hole spacetimes.
\vspace{15pt}
\subsection{Numerical Methods}
\label{sec:numerical_methods}
In order to solve the constraint and evolution equations from the previous sections, we use the numerical-relativity code \textsc{BAM}. While its functionality for one or two black holes in vacuum was developed in e.g.\ \cite{Brugmann1999, Brugmann2004, Brugmann2008}, a key step was the generalization to $N > 2$ in \cite{Galaviz2010}. For the present work, we had to perform several minor updates to re-enable more than two black holes after major changes to the code for general relativistic hydrodynamics \cite{Thierfelder2011, Dietrich2015, Schianchi:2023uky, Neuweiler:2024jae}. More recently, there have been significant improvements to the OpenMP parallelization of \textsc{BAM}. So while the basic $N>2$ functionality was already available in 2010, we ported/reimplemented certain features for the current, optimized version of \textsc{BAM}.

One aspect of the $N>2$ puncture method is that it is compatible with \textsc{BAM's} adaptive mesh refinement (AMR) algorithm. \textsc{BAM} uses ``nested moving boxes'', a Berger-Oliger type AMR in which the computational domain consists of a hierarchy of nested cell-centered Cartesian grids~\cite{BO, Brugmann1999}. If two or more of the moving cubical boxes overlap, they are replaced by their bounding box, which is the smallest rectangular box that contains all the original overlapping boxes. Generalizing from $N\leq 2$ to $N>2$ entails finding bounding boxes for more than two boxes. The grid hierarchy is set up such that every box on every level needs to be fully covered by a parent-level box on the next coarser level. Data is transferred between the individual levels via sixth-order Lagrangian polynomial interpolation, where the three-dimensional interpolant is obtained by successive one-dimensional interpolations.

The time evolution is based on the method-of-lines approach using fourth-order finite differencing in space and an explicit fourth-order Runge-Kutta time stepping. For the time evolution, we employ radiative boundary conditions, six buffer points for the transmission of gravitational waves through mesh-refinement boundaries, lop-sided advection stencils, quadratic interpolation in time, as well as Kreiss-Oliger dissipation of sixth order with a dissipation factor of 0.5 on all levels, except on level 0, where we choose a factor of 0.1. This improves stability and reduces high-frequency noise from mesh refinement boundaries. We track the puncture positions as described in~\cite{Campanelli2005,Brugmann2008} and compute the (coordinate) center of mass as the mass-weighted sum of the coordinate vectors.

Due to the chaotic nature and the resulting high sensitivity of $N$-black-hole evolutions to the initial conditions, accurate initial data are essential to obtain reliable results over reasonable time scales. In order to obtain the initial conformal factor $\psi$, we solve the Hamiltonian constraint (\ref{eq:ham_constraint_psi}) for the finite correction $u$ described in Equation (\ref{eq:psi_split}). For simplicity, we employ \textsc{BAM's} built-in multigrid solver to solve this nonlinear elliptic equation numerically with a second-order method, although in \cite{Galaviz2010} we also showed that up to 8th order multigrid is possible despite limited regularity at the punctures. As an aside, most $N=1$ or $N=2$ initial data solvers adapt coordinates and/or grid layouts to these special cases, see e.g.\ the two-puncture construction of \cite{Ansorg2004}.

We found that, contrary to simulations of (quasi\mbox{-)}circular black-hole binaries, a much lower threshold for the tolerance of the $l2$ norm of the residual is needed when computing initial data for general $N$-black-hole spacetimes. A tolerance of $10^{-10}$ led to reliable results with quasi-circular binaries in all of our test cases. On a few levels, this tolerance could not be guaranteed, and the convergence turned out to be very slow, so we stopped the computations after a maximum of 200 multigrid V-cycles on each level. With these initial data, the initial circularity of a binary in an $N$ black hole configuration is still not as good as in the case of a single (moving) binary, but good enough for our initial test cases.

In summary, the choice of the puncture method basically reduces the issue of going beyond two black holes to introducing loops over $N$ black holes wherever appropriate, as long as we do not encounter special constructions for $N=1$ or $N=2$. For example, this would not be possible for coordinates and grids tailored exclusively to two objects, say two excision spheres, etc.

However, in our work, we encountered the following additional challenges. First of all, there is the basic geometry of interesting $N=3$ or $N=4$ configurations, with by default much larger separations between the black holes than in typical binaries. In fact, a key limiting factor in current simulations are large separations, implying large boxes and additional demands for the grid sizes and computational costs.

Another comment concerns parallelization efficiency for more than two black holes, which however turned out to not be a major issue, assuming reasonable efficiency for $N=2$. BAM utilizes hybrid parallelization of MPI combined with OpenMP. The dynamic grid hierarchy with moving and varying boxes introduces a communication overhead. By design, the work for each box is parallelized separately without load balancing across several boxes. Initially, each puncture is assigned a single cubical box with the required finest grid spacing for a given puncture mass, and each box by itself represents a large work load with regard to parallelization. The algorithm works on the boxes sequentially, while being MPI and OpenMP parallelized for each individual box. Having more than two black holes did not introduce additional complications here, requiring computations scaling approximately linearly with the number of boxes. However, the large-separation configurations with multiple boxes can lead to more dynamic box configurations, with more frequent regridding operations for merging and unmerging boxes. In the present simulations, this does not introduce a major overhead but could become an issue for more ambitious simulations.

The gravitational wave extraction is performed by using the Newman-Penrose scalar $\Psi_4$ as described in \cite{Brugmann2008}. A fundamental issue arises for configurations with more than two black holes. We typically require a reference frame with a center and an axis to define a tetrad basis. There are well-known issues related to the choice of frame, leading to various effects such as mode mixing.

The accuracy of our numerical simulations is evaluated in Appendix \ref{sec:numerical_accuracy}. In particular the constraint violations, the conservation of mass, linear momentum and angular momentum as well as the behavior of the gravitational-wave signals at different extraction radii are discussed.

\subsection{Post-Newtonian Methods}
\label{sec:pn_methods}

For the post-Newtonian (PN) approximations of the binary-single encounters up to 2.5PN we use the three-body Hamiltonian and the equations of motion presented in \cite{Schaefer1987, Lousto2008b, Galaviz2011, Galaviz2011b, Bonetti2016, Zevin2019} and we performed similar tests to verify the validity of our results. We integrate the equations of motion using a 5th order Runge-Kutta scheme with an adaptive step size to keep the relative error below $10^{-6}$.

In the context of $N$ black holes, let us point out that it is precisely the case $N=2$ that can be treated in great detail since it allows the reduction to an effective one-body problem in the Newtonian, and more importantly also in the post-Newtonian case to various orders. There are comparatively very few papers on $N=3$, and there even exist obstructions to generalize the $N=3$ treatment of \cite{Schaefer1987} to $N>3$ without further assumptions. We therefore limit our PN computations to $N\leq3$ and to order 2.5PN.

\section{Configuration Setup}
\label{sec:exp_setup}
We set up the black-hole binary-single and binary-binary scattering experiments in a way similar to how it has been done in \cite{Damour2014} for the scattering of two black holes. This ensures an efficient use of the available computational domain, with the scattering objects having a large initial separation.

\begin{figure}[htp!]
    \centering
    \includegraphics[width=1.0\linewidth]{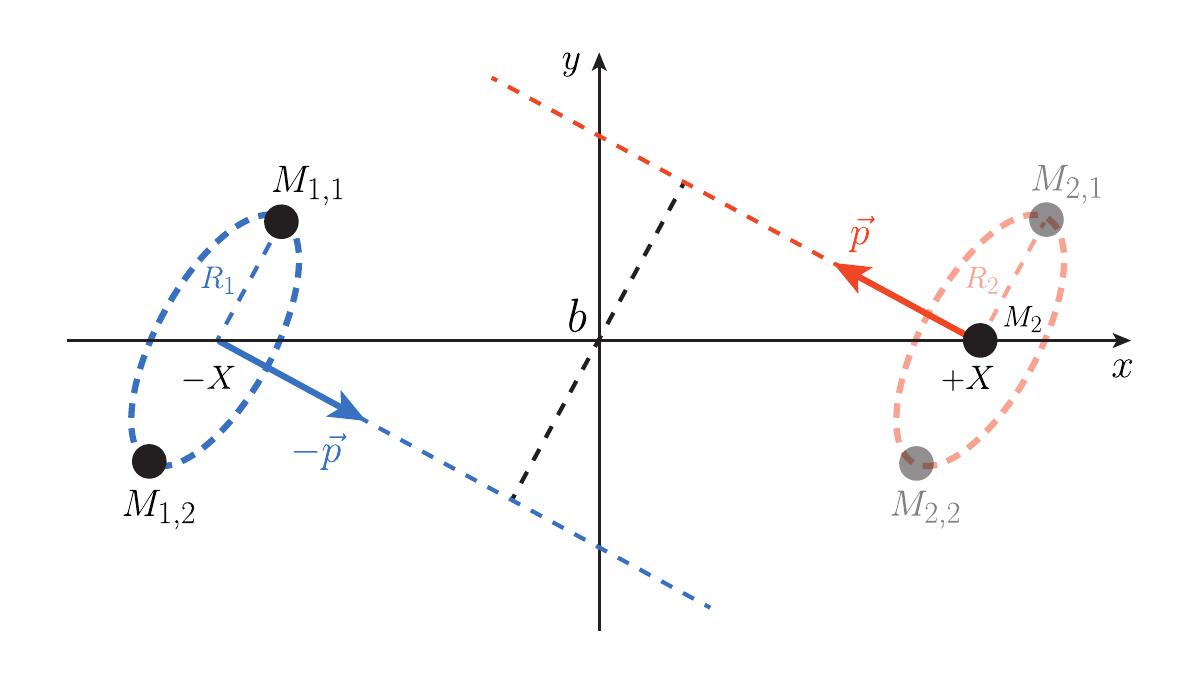}
    \caption{Schematic showing the configuration setup of the black-hole binary-single and binary-binary encounters.}
    \label{fig:experimental_setup}
\end{figure}

Figure \ref{fig:experimental_setup} depicts the setup schematically. Both the black-hole binary's center of mass and the single black hole (or the other black-hole binary's center of mass) start on the $x$-axis at locations $\pm X$, resulting in an initial separation of $2X$. The momenta $\pm \vec{p}$ lie in the $xy$-plane and have components
\begin{equation}
    \pm \vec{p} = \pm |\vec{p}| \begin{pmatrix}-\sqrt{1-\left(\frac{b}{2X}\right)^2} \\ \frac{b}{2X} \\ 0\end{pmatrix},
\end{equation}
which leads to an encounter with an impact parameter $b$. The orientation of the binary's initial orbital plane can be chosen arbitrarily. In some of the numerical experiments, we rotate the entire configuration such that the orbital angular momentum vector of the initial binary (or binaries) is (anti-)parallel to the $z$-axis, thereby aligning it with the tetrad basis used in the construction of $\Psi_4$.

The complete parameter space of binary-single scattering events is 9-dimensional, with the relevant parameters being the mass ratio of the single black hole and the binary, the mass ratio of the black holes in the binary, the binary's initial eccentricity, its initial orientation (which can be described by three orientation angles), the phase of the binary at the time when the incoming single black hole has reached its closest distance to the binary's center of mass, the impact parameter $b$ expressed in units of the initial semi-major axis of the binary, and the magnitude of the relative approach velocity at infinity $v_\infty$. For binary-binary scattering experiments, one additionally has to specify the mass ratio, the initial semi-major axis, eccentricity, orientation, and phase of the second binary, leading to a total of 16 relevant parameters. 

For a first exploration of these vast and high-dimensional parameter spaces, we restrict ourselves to the following initial configurations: We set the masses of all black holes to be equal, and for all black hole binaries we use initial parameters from \cite{Tichy2011} with reduced eccentricity and an initial coordinate separation of 11.9718 M. We add the momenta from these binary black hole initial parameters to the additional motion of the binary coming from the momentum $\vec{p}$. Depending on the magnitude of $\vec{p}$ we adjusted the puncture masses such that the ADM mass at each puncture is always equal to 0.5 with deviations that do not exceed 0.1\%. We further restrict ourselves to a small subset of possible binary orientations: one set where the normal vector of the orbital plane is (anti-)parallel to the initial momentum and one where the normal vector is orthogonal to the plane in which the binary's center of mass and the single black hole initially move.

At this point, we want to highlight that due to the hierarchical grid setup described in Section \ref{sec:methods}, the boxes moving with the black holes have to be inside the smallest non-moving box at all times. For each grid setup, this sets a limit to the maximum initial distance $X_{\mathrm{max}}$ of the scattering objects and to the time up to which the simulation can be evolved. As soon as one of the black holes is about to leave the smallest non-moving box, the simulation stops immediately, which can become a very limiting factor, not only for studying the long-term dynamics of the system at hand but also for extracting the gravitational-wave signal at sufficiently large radii. Especially for high relative initial velocities of the scattering objects, the gravitational-wave signal from the encounter might not be able to travel far enough until one of the black holes leaves the available domain.

We use three different grid setups, depending on the required size of the domain in which the black holes should be able to move. In all of our simulations, the grid hierarchy contains 11 levels indexed by $L=0, 1, ..., 10$. Levels 0 to 5 consist of fixed cubical boxes with $n$ grid points in each direction and a grid spacing of $h_L = h_0/2^L$, that are centered at the origin. Levels 6 to 10 consist of moving cubical boxes that are centered at the locations of the punctures, with $n/2$ grid points in each direction and a grid spacing of $h_L = h_0/2^L$. Our values for $n$ and $h_0$ for the three different grid setups can be found in Tables \ref{tb:grid_1} to \ref{tb:grid_3}. In some cases, the number of grid points is slightly adjusted to guarantee a total amount of $4n+2$ grid points in each direction. \\

\begin{table}[htp!]
\begin{tabular}{l|l|l|l|l|l}
\hline
\multicolumn{1}{|l|}{\textbf{$n$}} & 112 & 134 & 168 & 224 & \multicolumn{1}{l|}{336} \\ \hline
\multicolumn{1}{|l|}{\textbf{$h_0$}} & 24 & 20.0597 & 16 & 12 & \multicolumn{1}{l|}{8} \\ \hline
\multicolumn{1}{|l|}{\textbf{$h_{10}$}} & 0.02344 & 0.01959 & 0.01563 & 0.01172 & \multicolumn{1}{l|}{0.00781}  \\ \hline
\end{tabular}
\caption{Grid setups for a maximum black-hole $x$-coordinate of $X_{\mathrm{max}} \approx 61M$ (the same goes for the $y$- and $z$-coordinate) and a domain with a cubical side length of $l=2688M$ .}
\label{tb:grid_1}
\end{table}

\begin{table}[htp!]
\begin{tabular}{l|l|l|l|l|l}
\hline
\multicolumn{1}{|l|}{\textbf{$n$}} & 144 & 172 & 216 & 288 & \multicolumn{1}{l|}{432} \\ \hline
\multicolumn{1}{|l|}{\textbf{$h_0$}} & 24 & 20.0930 & 16 & 12 & \multicolumn{1}{l|}{8} \\ \hline
\multicolumn{1}{|l|}{\textbf{$h_{10}$}} & 0.02344 & 0.01953 & 0.01563 & 0.01172 & \multicolumn{1}{l|}{0.00781}  \\ \hline
\end{tabular}
\caption{Grid setups for a maximum black-hole $x$-coordinate of $X_{\mathrm{max}} \approx 83M$ (the same goes for the $y$- and $z$-coordinate) and a domain with a cubical side length of $l=3456M$.}
\label{tb:grid_2}
\end{table}
\vspace{-15pt}
\begin{table}[htp!]
\begin{tabular}{l|l|l|l|l|l}
\hline
\multicolumn{1}{|l|}{\textbf{$n$}} & 192 & 230 & 288 & 384 & \multicolumn{1}{l|}{576} \\ \hline
\multicolumn{1}{|l|}{\textbf{$h_0$}} & 24 & 20.0348 & 16 & 12 & \multicolumn{1}{l|}{8} \\ \hline
\multicolumn{1}{|l|}{\textbf{$h_{10}$}} & 0.02344 & 0.01953 & 0.01563 & 0.01172 & \multicolumn{1}{l|}{0.00781}  \\ \hline
\end{tabular}
\caption{Grid setups for a maximum black-hole $x$-coordinate of $X_{\mathrm{max}} \approx 110M$ (the same goes for the $y$- and $z$-coordinate) and a domain with a cubical side length of $l=4608M$.}
\label{tb:grid_3}
\end{table}
\vspace{-15pt}
\begin{figure*}
\centering
\includegraphics[width=\textwidth]{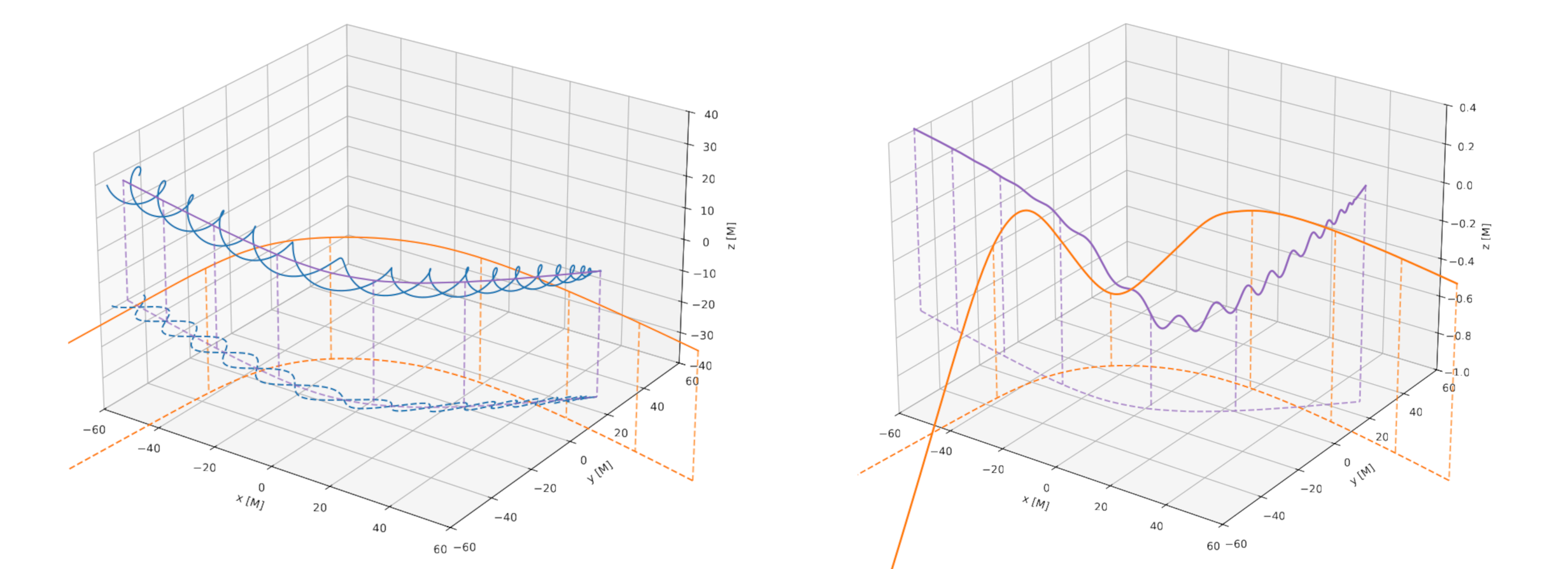}
\caption{\textit{Left:} Three-quarter view of the puncture trajectories of BSS-01a. \textit{Right:} The same view but without the trajectories of the black holes in the binary and with a strongly rescaled $z$-axis. The trajectory of the single black hole (coming from the right) is colored orange, the trajectory of the center of mass of the binary system (coming from the left) is colored purple, and the trajectories of the binary members are colored blue. The dashed lines and projections indicate the location in 3D space.}
\label{fig:bs_scattering02_trajectories}
\end{figure*}

\section{Results}
\label{sec:results}

\subsection{Weak Binary-Single Scattering (BSS-01)}
\label{sec:weak_bss}
Our first case study is a black-hole binary-single scattering with $X=100M$ (BSS-01a) and an impact parameter $b=50M$ that is several times greater than the initial coordinate separation of the black holes that make up the binary, leading to a relatively weak perturbation of the binary dynamics. The binary is oriented such that the normal vector of the orbital plane is parallel to the momentum vector $-\vec{p}$, i.e.\ the orbital plane of the binary is chosen to be approximately perpendicular to the scattering motion occurring mostly in the $xy$-plane. The initial relative approach velocity of the binary and the single black hole is 0.223. A detailed list of all the initial parameters can be found in Table \ref{tb:parameter_values} in Appendix \ref{sec:extra_material}. 

\begin{figure}[htp!]
    \centering
    \includegraphics[width=1.0\linewidth]{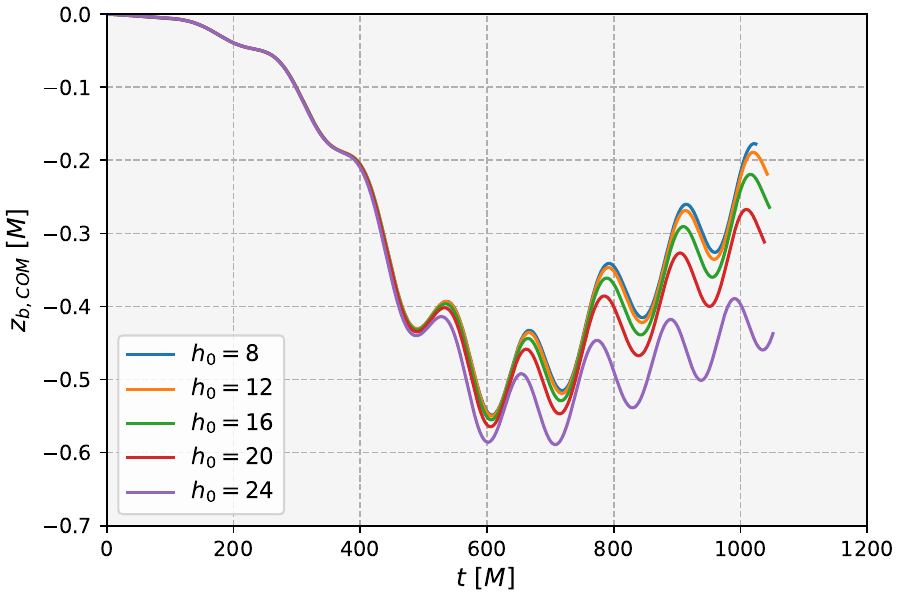}
    \caption{Convergence test for the $z$-component of the binary's center of mass for the configuration BSS-01b.}
    \label{fig:convergence_test}
\end{figure}

\begin{figure*}[htp!]
\includegraphics[height=0.88\columnwidth]{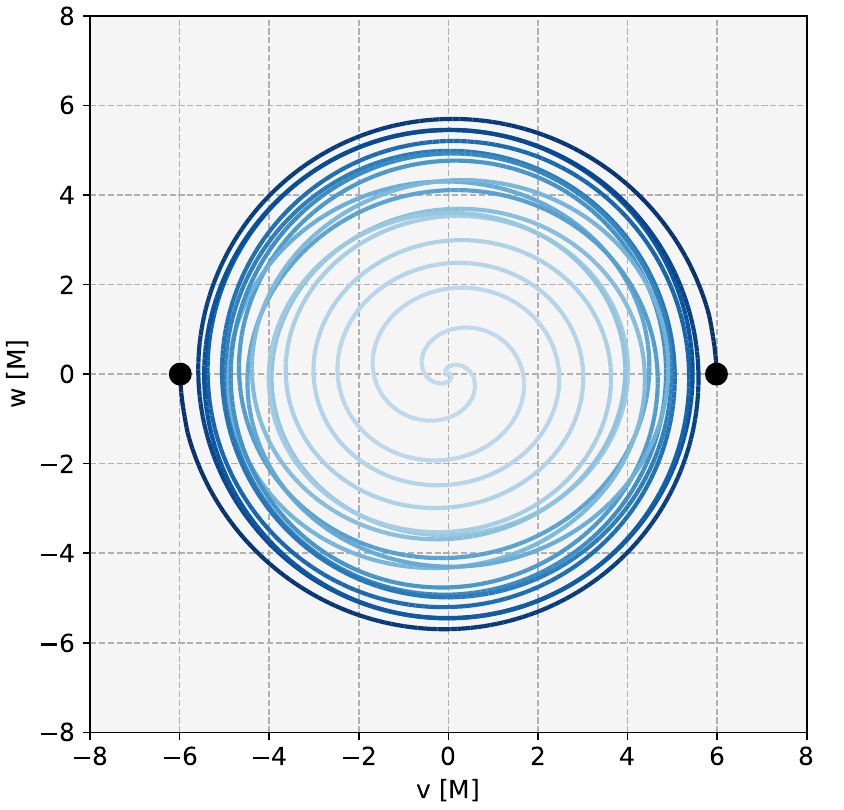}
\hspace{8pt}
\includegraphics[height=0.88\columnwidth]{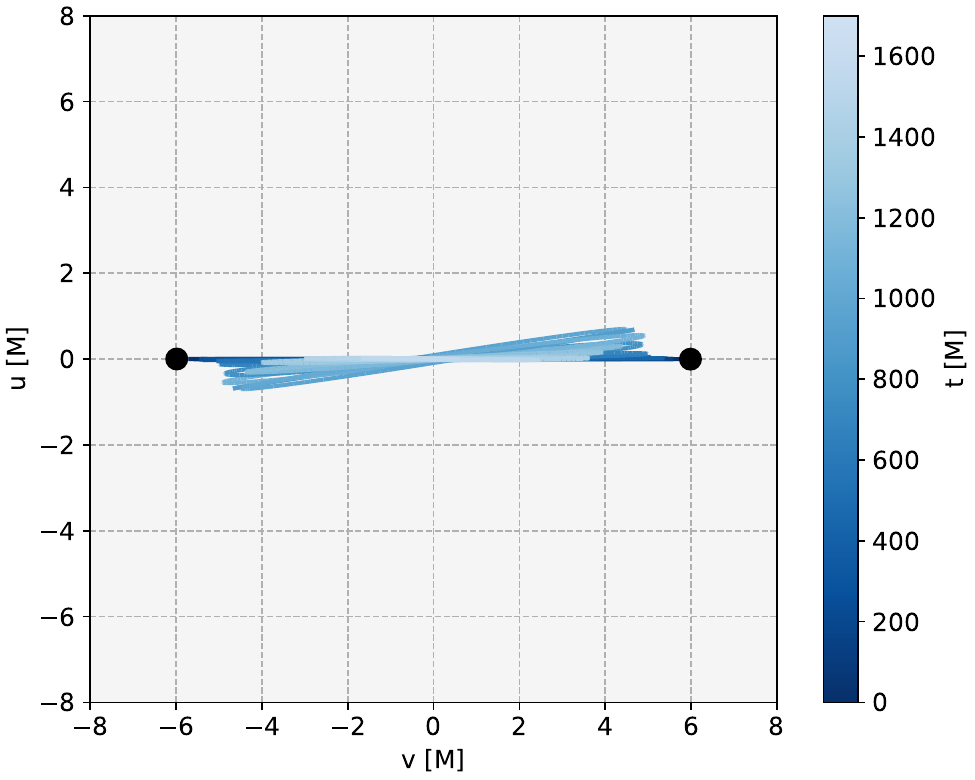}
\caption{Puncture trajectories of the binary members in the setup BSS-01a in their center-of-mass frame, which uses coordinates $u$, $v$ and $w$. The color encodes the positions of the black holes at different times, and the initial positions of the black holes are indicated with black dots. \textit{Left:} Head-on view of the initial orbital plane. \textit{Right:} Edge-on view of the initial orbital plane.}
\label{fig:bs_scattering02_trajectories_com}
\end{figure*}
\begin{figure}
    \centering
    \hspace{-22pt}
    \includegraphics[width=0.94\columnwidth]{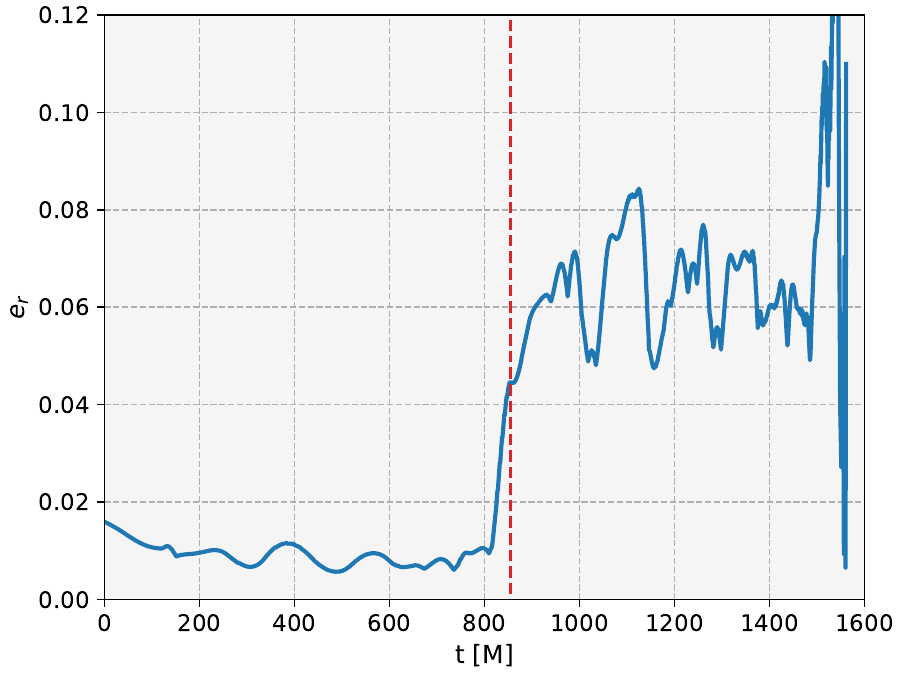}
    \caption{Time evolution of the binary's eccentricity $e_{\mathrm{r}}$ in BSS-01a. The dashed red line marks the time of closest approach of the binary's center of mass and the single black hole.}
    \label{fig:bs_scattering02_ecc_evolution}
\end{figure}

Figure \ref{fig:bs_scattering02_trajectories} shows two three-quarter views of the trajectories of the black holes, where in the right plot the $z$-axis is strongly rescaled to show the motion in the $z$-direction. One can see noticeable but small (1\% of the binary separation) oscillations of the binary's center of mass, which have a frequency approximately equal to twice the binary's orbital frequency and an amplitude that slowly increases and decreases as the single black hole is approaching and receding away again. Some amount of acceleration in the $z$-direction has to be expected since the configuration is not planar, with the binary moving outside the $xy$-plane, and the distance between the single black hole and the binary constituents being not symmetric. Therefore, part of the effect is already Newtonian, plus additional effects due to the coupling of the orbital angular momentum of the binary to the center-of-mass/single orbit, similar to spin-orbit coupling.

We performed a convergence study using a similar setup with the same impact parameter and initial phase of the binary, but for $X=50M$ (BSS-01b). We used all the resolutions in Table \ref{tb:grid_1} and for late times the solutions show monotonic convergence to an oscillating solution. Figure \ref{fig:convergence_test} shows the $z$-component of these oscillations.

Figure \ref{fig:bs_scattering02_trajectories_com} shows the puncture trajectories of the binary black holes in their center-of-mass frame which, is described by the new coordinates $u$, $v$ and $w$ (in this frame the initial orbital plane lies in the $vw$-plane). One can see that the influence of the single black hole causes the orbital plane to rotate by approximately 10 degrees. Towards the merger, the orbital plane rotates back to its initial orientation. Another important effect of the single black hole is an increase in the eccentricity of the binary, which can be seen qualitatively in Figure \ref{fig:bs_scattering02_trajectories_com} on the left, but also quantitatively in Figure \ref{fig:bs_scattering02_ecc_evolution}. We define the eccentricity $e_{\mathrm{r}}$ as
\begin{equation}
    e_{\mathrm{r}}(t) = \frac{\Delta r_{\mathrm{max}}(t) - \Delta r_{\mathrm{min}}(t)}{2 r_{\mathrm{avg}}(t)},
\end{equation}
where the average separation, and the extremal deviations from a smoothed value $r_{\mathrm{s}}$ are given by
\begin{align}
    r_{\mathrm{avg}}(t) &= \frac{1}{T} \int^{t+T/2}_{t-T/2} dt' r(t'), \\
    \Delta r_{\mathrm{max}}(t) &= \max_{t' \in [t-T/2, \ t+T/2]} [r(t') - r_{\mathrm{s}}(t',t)], \\
    \Delta r_{\mathrm{min}}(t) &= \min_{t' \in [t-T/2, \ t+T/2]} [r(t') - r_{\mathrm{s}}(t',t)],
\end{align}
and where the period $T$ is defined using Kepler's law
\begin{equation}
    T = 2 \pi (r^3/M)^{1/2}.
\end{equation}
Here $M$ is the combined mass of the black holes in the binary. The smoothed value $r_s(t',t)$ is obtained from
\begin{equation}
    r_{\mathrm{s}}(t', t) = r(t) + \frac{r(t+T/2) - r(t-T/2)}{T} (t'-t)
\end{equation}
(for more details and comparisons to other definitions of the eccentricity see, e.g., \cite{Tichy2011, Ficarra2025}). \\
\begin{figure*}[htp!]
\includegraphics[width=1.0\textwidth]{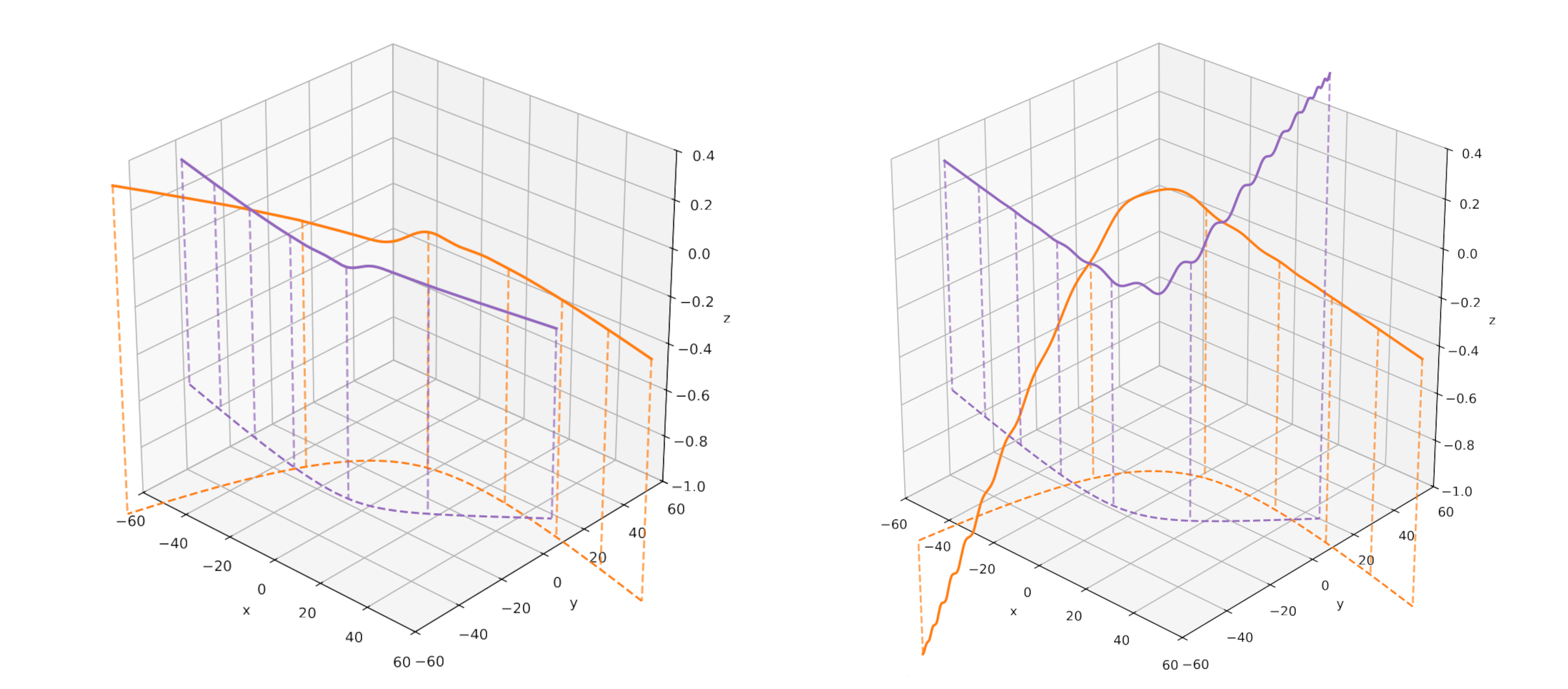}
\caption{Three-quarter views of the binary-single scattering puncture trajectories in the Newtonian case (on the left) and in the post-Newtonian case with correction terms up to 2.5PN (on the right), using the same initial parameters as in Figure \ref{fig:bs_scattering02_trajectories}. The trajectory of the binary's center of mass is colored purple, and the trajectory of the single black hole is colored orange.}
\label{fig:pn_trajectories01}
\end{figure*}
\begin{figure}
    \centering
    \includegraphics[width=0.94\columnwidth]{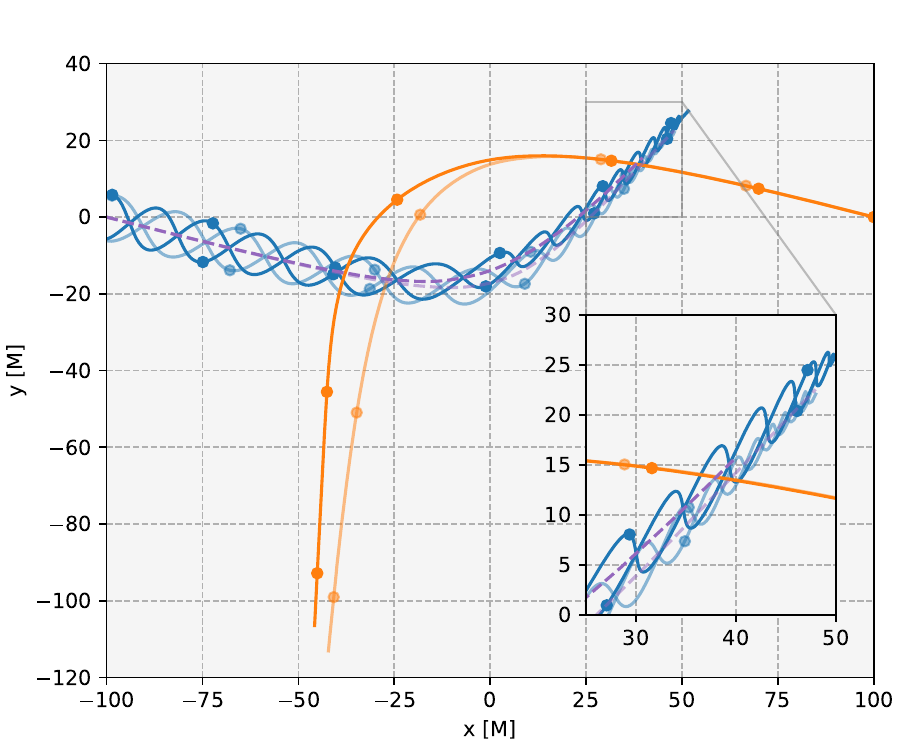}
    \caption{Comparison of the puncture trajectories (BSS-01a) of the fully relativistic simulation (fully opaque) and the post-Newtonian simulation up to order 2.5PN (transparent) using the same initial parameters. The positions of the black holes are shown for different time steps separated by 300 M.}
    \label{fig:nr_25pn_comparison}
\end{figure}
In Figure \ref{fig:bs_scattering02_ecc_evolution} one can observe that the eccentricity increases at a relatively high rate around the time of closest approach of the binary's center of mass and the single black hole (dashed red line). After the closest approach, the eccentricity oscillates around a value of roughly 0.07, and the amplitude of these oscillations decreases over time. Before the eccentricity oscillations can fully decay, the plunge and merger occur. We want to note that some of the observed effects could be partially associated with a non-zero spin of the black holes in the binary after the encounter. As described earlier, the black holes all start without spin, but scattering processes can lead to the spin-up of initially non-spinning black holes \cite{Nelson2019, Jaraba2021}. We plan to address the analysis of spins in binary-single and binary-binary encounters in a future study. 

We finally compare the dynamics of the fully relativistic case with those obtained for an analogous system in successive post-Newtonian approximations. Figure \ref{fig:pn_trajectories01} shows the trajectories of the binary's center of mass and of the single black hole in the Newtonian case (on the left) and when using all the post-Newtonian correction terms up to order 2.5PN (on the right). In the Newtonian case, the motion in the $z$-direction is relatively small, and the scattering objects get deflected into different directions compared to the 2.5PN and the fully relativistic case. In the 2.5PN case, there are oscillations similar to those in the fully relativistic case (see e.g.\ Figure \ref{fig:convergence_test}). These oscillations already occur at the 1PN order, and they do not decay if one does not employ the dissipative effects of the 2.5PN order. In the fully relativistic case, the single black hole does not show such oscillations, and the overall small-scale motion in the $z$-direction looks quite different from the 2.5PN approximation. Figure \ref{fig:nr_25pn_comparison} shows the direct comparison of the fully relativistic and the 2.5PN simulation in the $xy$-plane using the same coordinate values. The trajectories look surprisingly similar, even though the coordinates used in both cases are different. The strongest deviations occur close to the point of closest approach. The orbital decay rate is much higher in the post-Newtonian approximation, and a significant phase shift is introduced over time. In this setup, the dynamics do not depend strongly on the initial phase of the binary in all cases (Newtonian, post-Newtonian, and fully relativistic). Only the orientation and direction of rotation of the binary determine which object moves upwards and which object moves downwards. 

\begin{figure*}[htp!]
\includegraphics[width=0.85\columnwidth]{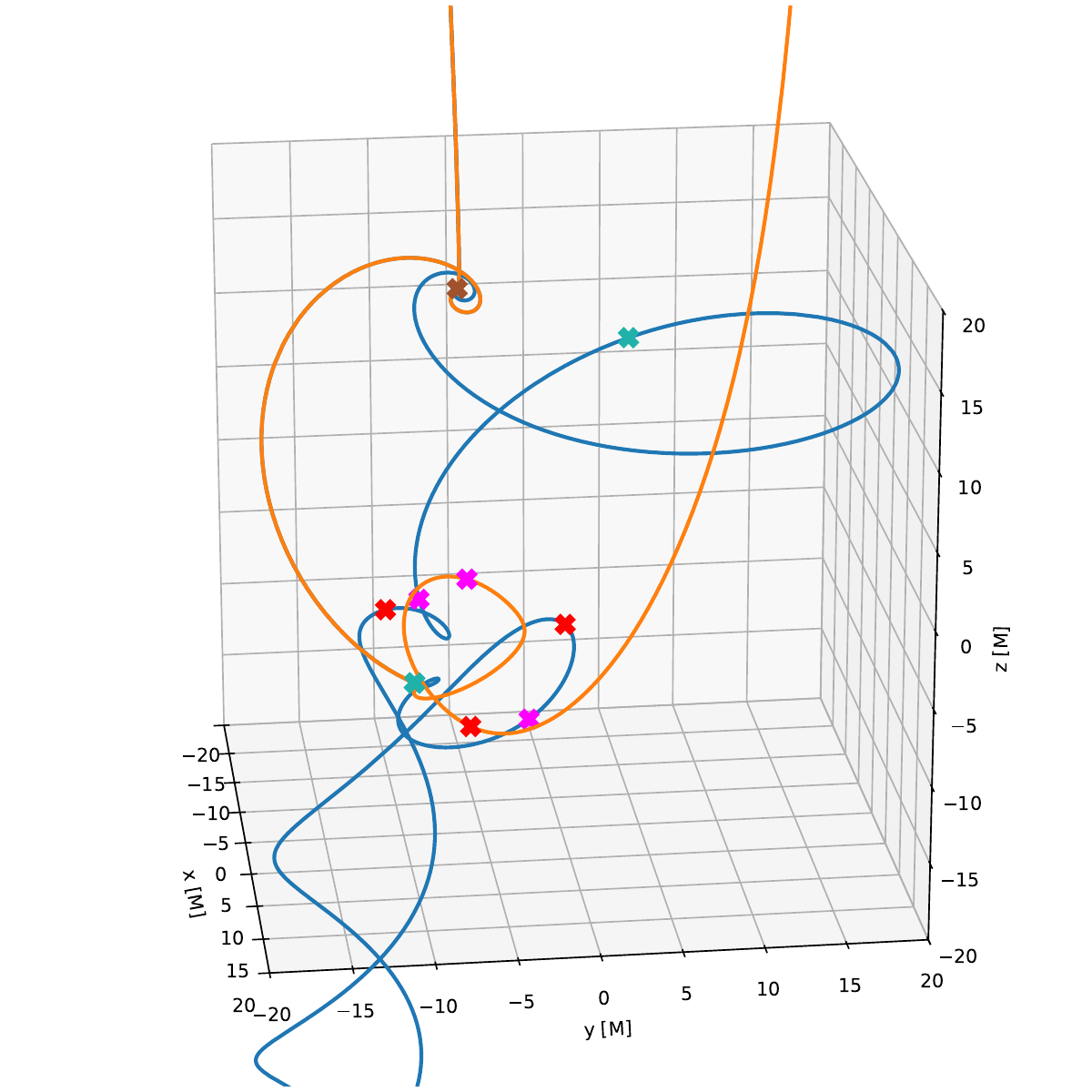}
\hspace{10pt}
\vspace{10pt}
\includegraphics[width=1.05\columnwidth]{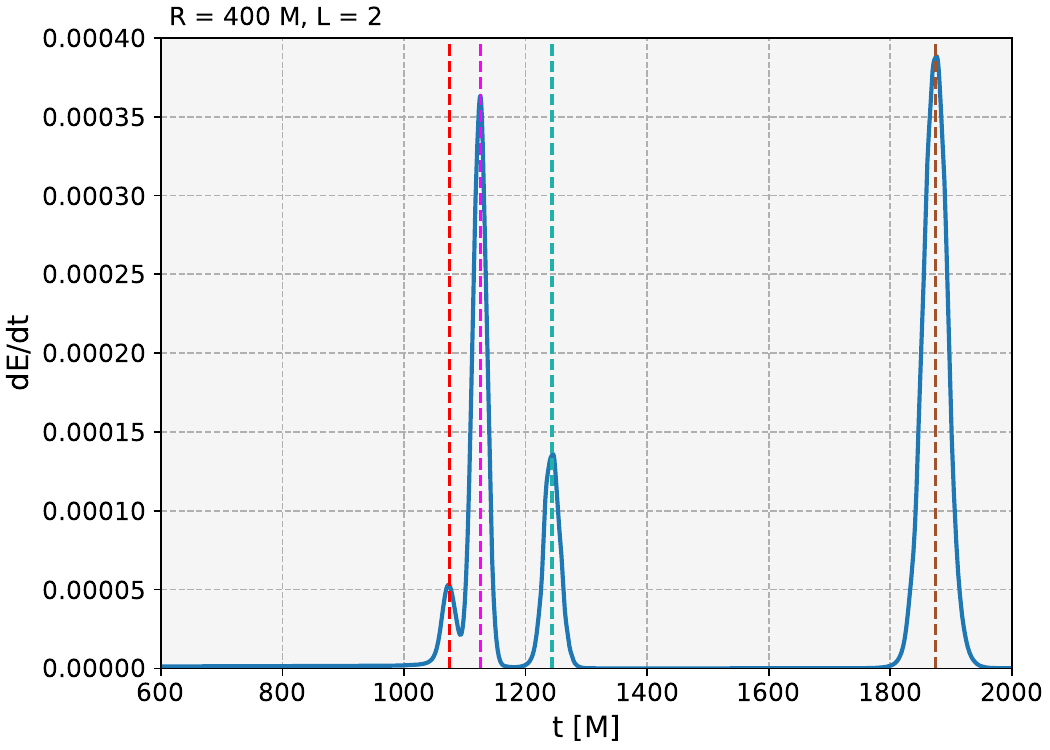}
\includegraphics[width=0.95\textwidth]{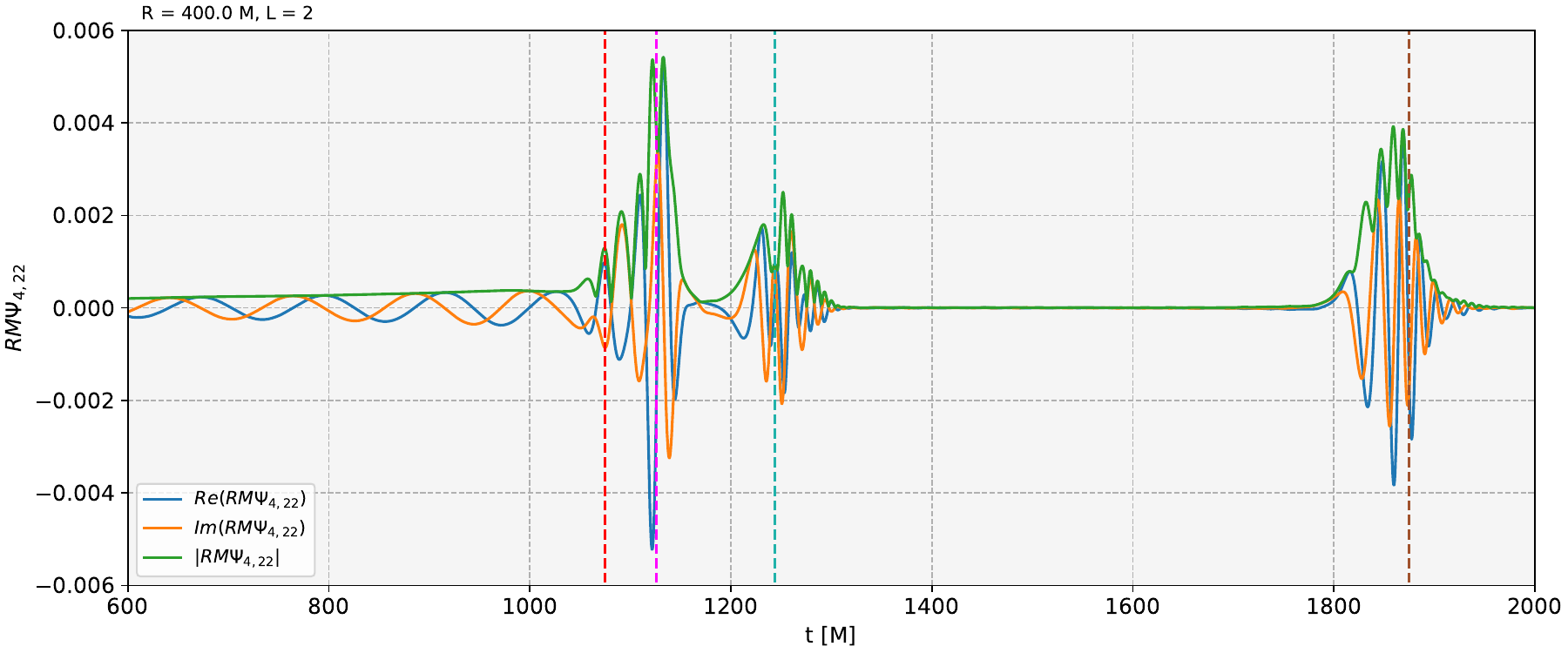}
\caption{\textit{Top left:} Puncture trajectories of the black holes of BSS-02 (colored blue for the black holes in the binary and orange for the single black hole). The colored crosses correspond to the times marked by the dashed lines in the other plots. \textit{Top right:} Gravitational-wave luminosity extracted at the radius $R=400M$ and on the level $L=2$. \textit{Bottom:} The real part (blue), the imaginary part (orange), and the absolute value (green) of the 22-mode of $RM\Psi_4$, extracted at the same radius and level.}
\label{fig:bss15c_rpsi4m22}
\end{figure*}

\subsection{Complex Interaction With a Double Merger After a Strong Binary-Single Encounter (BSS-02)}
\label{sec:complex_interaction}

Here we present another setup with a close binary-single encounter resulting in complex interactions with multiple close passages and a double merger. For this we choose $X=75.0M$, $b = 30.0M$ and $|\vec{p}| = 0.05M$. The system of this section is rotated with respect to the setup described in Section \ref{sec:exp_setup} such that the orbital angular momentum vector of the initial binary is aligned with the $z$-axis and therefore also with the tetrad basis used to construct $\Psi_4$. The detailed list of initial parameters can be found in Table \ref{tb:parameter_values} in Appendix \ref{sec:extra_material}.

\begin{figure*}
    \centering
    \includegraphics[width=0.96\linewidth]{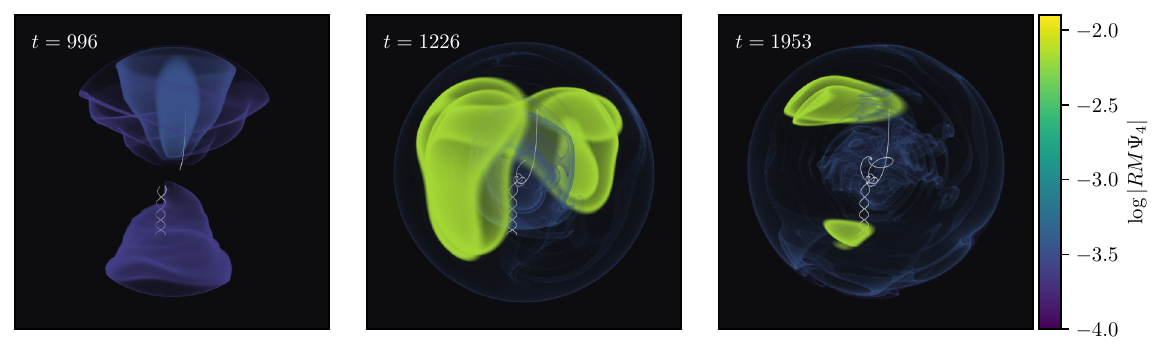}
    \caption{The $zy$-plane view on three-dimensional volume rendering of the amplitude of $|RM\Psi_4|$ of BSS-02 at different times and limited to the emission sphere of radius $R=200M$. The data is shown only at the isosurfaces of selected values of the amplitude to highlight the emission features, i.e., other amplitude values also contain emission. The puncture trajectories of the black holes are indicated in white. The times are picked to capture the emission from the moving binary ($t=996M$), the first two bursts of emission after encounter with the incoming black hole ($t=1226M$), and the last burst of emission from the coalescence of the newly formed binary ($t=1952M$).
    }
    \label{fig:3d_gw_vis}
\end{figure*}

Figure \ref{fig:bss15c_rpsi4m22} shows the puncture trajectories of the three black holes (top left), the gravitational-wave luminosity (top right), and the 22-mode of $RM\Psi_4$, extracted at the radius $R=400M$ and on the level $L=2$. The plot of the gravitational-wave luminosity shows four distinct peaks, which we marked with differently colored dashed lines and which correspond to the dashed lines in the plot for $RM\Psi_{4,22}$. One can connect the peaks in the gravitational-wave emission to specific events in the binary-single encounter. We illustrate this connection by adding crosses to the puncture tracks at the retarded times $t_R = t - R$, which approximately correspond to the moments in time $t$ marked by the dashed lines. The fact that the interactions do not happen at the origin of the grid and the space is not flat makes this a rough estimate, but it should allow for a qualitative assessment. The first peak (colored red) is connected with the first strong encounter, where the single black hole pulls both black holes in the binary in a different direction. The second peak (colored magenta) is associated with the second interaction of the single black hole, primarily with one of the binary members, which temporarily ejects it from the strong interaction region. The third peak (colored turquoise) is created at the merger of the single black hole with one of the binary members, and the fourth peak (colored brown) corresponds to the merger of the remaining black holes. 

\begin{figure*}[htp!]
\includegraphics[width=1.0\textwidth]{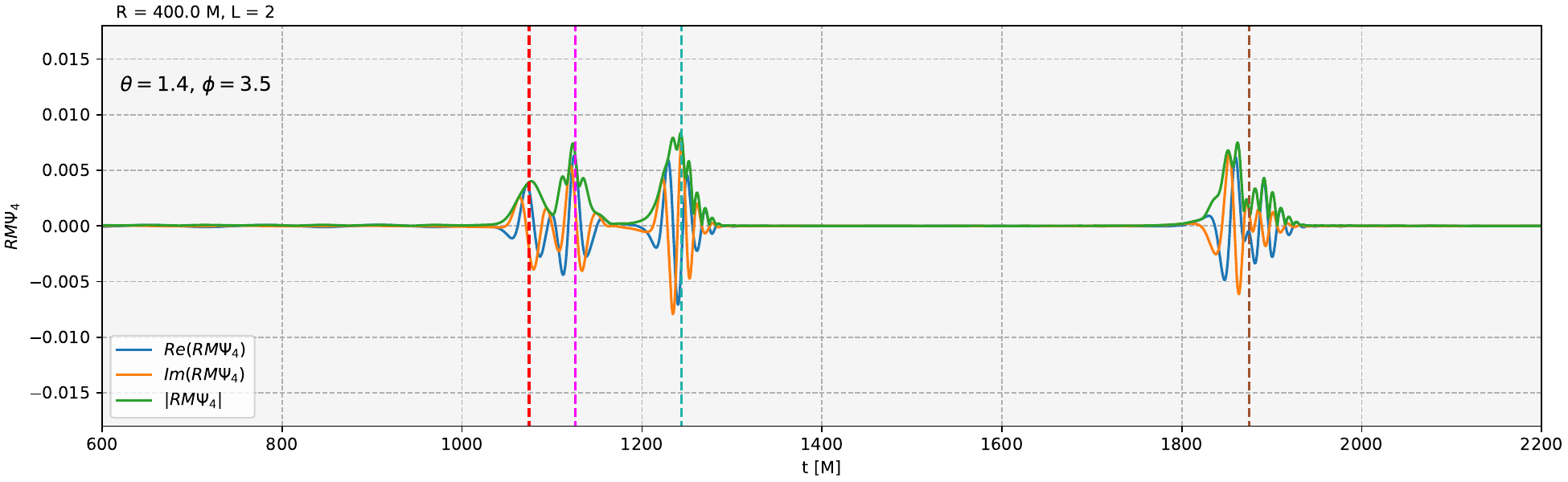}
\includegraphics[width=1.0\textwidth]{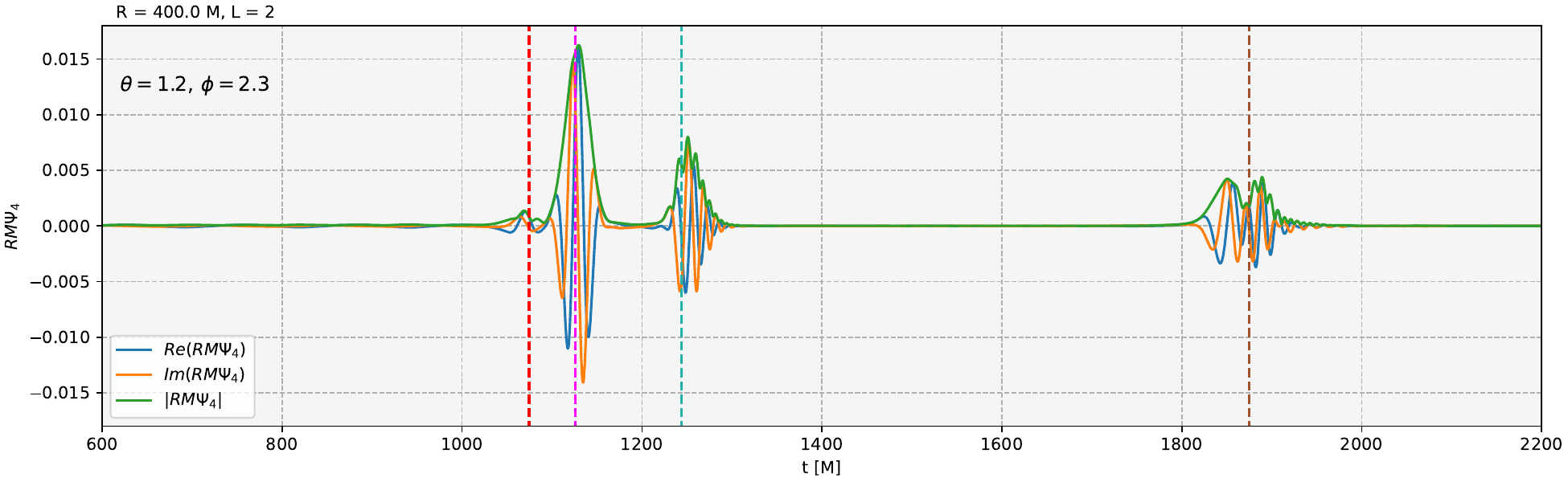}
\includegraphics[width=1.0\textwidth]{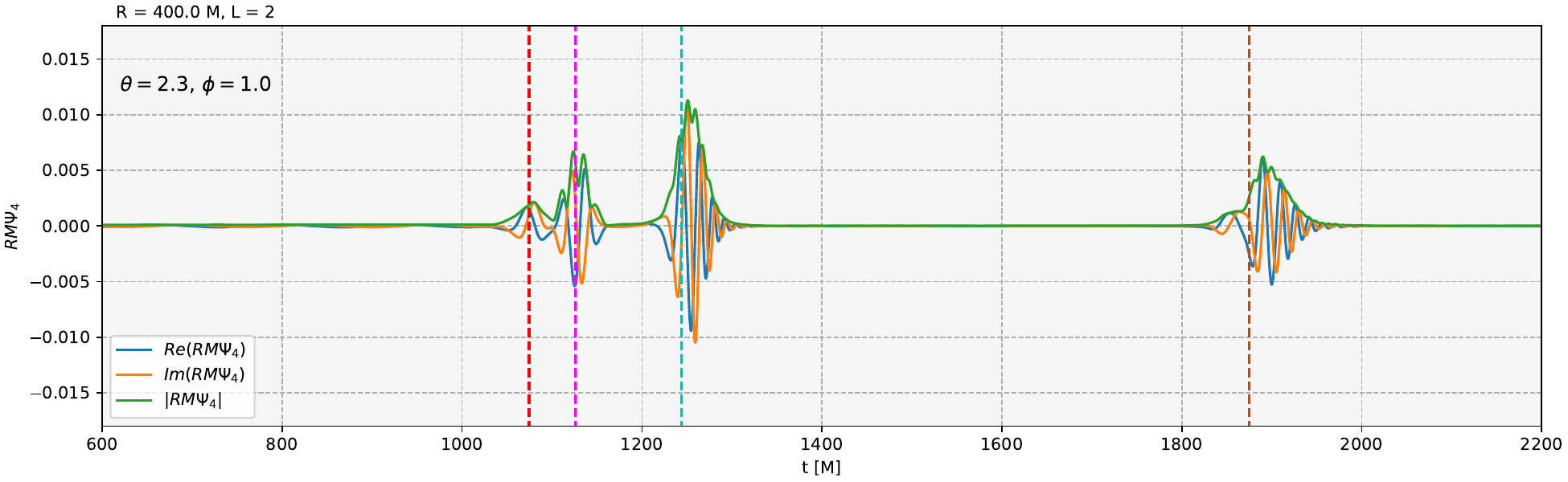}
\includegraphics[width=1.0\textwidth]{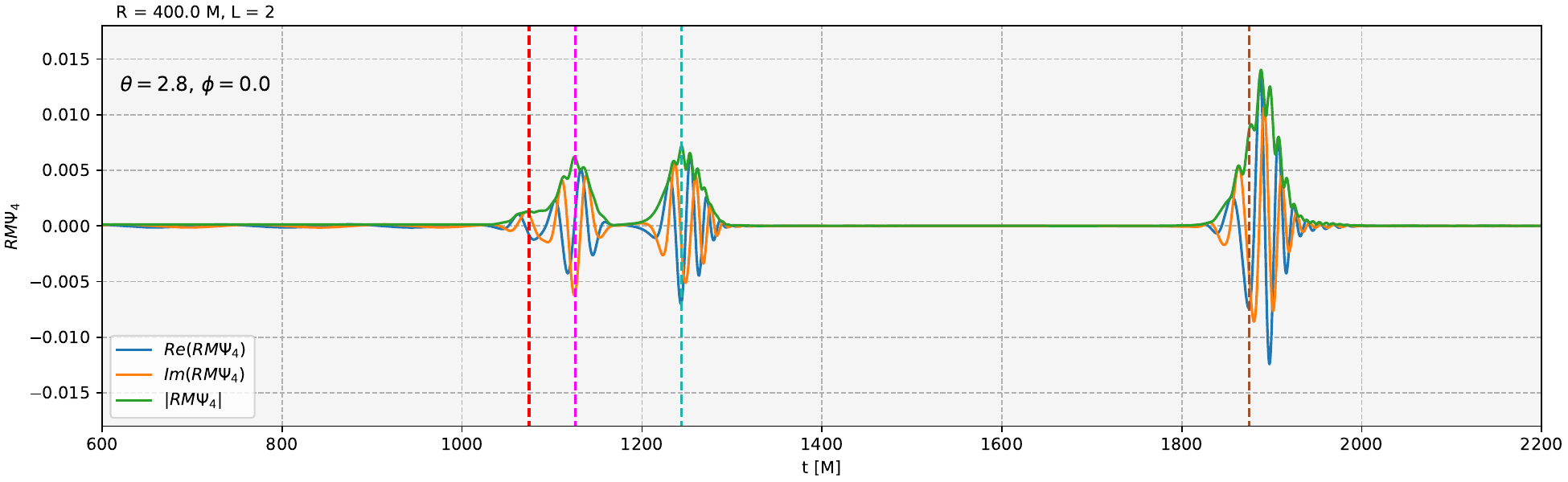}
\caption{The real part (blue), the imaginary part (orange), and the absolute value (green) of $RM\Psi_4$ for the system BSS-02 extracted at the radius $R=400M$ and on the level $L=2$. $RM\Psi_4$ is reconstructed from all modes from $l=2$ to $l=4$ and is evaluated in different directions (indicated at the top left). The colored dashed lines correspond to those in Figure \ref{fig:bss15c_rpsi4m22}.}
\label{fig:bss15c_rpsi4_directions}
\end{figure*}

We now discuss the gravitational-wave emission and extraction in more detail for this example. The 22-mode of $RM\Psi_4$ starts with a regular gravitational-wave signal of a quasi-circular binary and then shows several peaks that are associated with the peaks in the gravitational-wave luminosity. The first two peaks cannot be distinguished in the 22-mode, and all visible peaks show high-frequency oscillations in the absolute value of $RM\Psi_{4,22}$. We suggest that this is not a physical effect but rather a result of the source being displaced from the origin of the grid as well as moving, rotating, and accelerating within the extraction sphere, which causes mode mixing \cite{Boyle2016, Chaurasia2018, Woodford2019}, anisotropic signal arrival times on the extraction sphere across different directions, blueshifting of the signals in the direction of motion, and redshifting in the opposite direction. For comparison, we provide the 22-mode and the 44-mode of a merger signal of a black-hole binary that is moving uniformly with respect to the origin of the coordinate grid in Figure \ref{fig:gw_bbh_moving} in Appendix \ref{sec:extra_material}. The gravitational-wave luminosity, which is computed by using the total $\Psi_4$, which contains all the modes, does not show these spurious oscillations. Furthermore, the emission of gravitational waves in $N$-body systems can originate from very different locations and tends to be highly anisotropic, which is illustrated in Figure \ref{fig:3d_gw_vis}. The emissions from the initial binary are primarily in the $z$-direction (see the image on the left). In contrast, the first three bursts are sent in distinct directions that are mostly perpendicular to the $z$-axis. The emission of the first two bursts is depicted in the center panel of Figure \ref{fig:3d_gw_vis}. The last burst is primarily sent in the positive and negative $z$-direction (see the image on the right). Because of this and the mode mixing, the 22-mode is not necessarily the most dominant and representative mode. As an alternative, one can reconstruct $RM\Psi_4$ on the extraction sphere using all of the available modes and evaluate it in different directions. This is illustrated in Figure \ref{fig:bss15c_rpsi4_directions}, which shows $RM\Psi_4$ reconstructed from all modes from $l=2$ to $l=4$ for four different directions. The directions were picked such that each direction maximizes the amplitude of the absolute value of $RM\Psi_4$ for a given peak. The spurious oscillations in the absolute value are still present but slightly weaker than in the 22-mode. Due to the mode mixing, modes with higher $l$ also need to be considered and are no longer subdominant. We tried using all the modes up to $l=6$ which did not significantly reduce the oscillations but slightly changed the amplitudes of some peaks. When evaluating the full $\Psi_4$ computed at each grid point in the simulation across various locations on the extraction sphere, oscillations in $|RM\Psi_4|$ persist, suggesting that mode mixing and considering only a finite number of modes is not the only contributor to these oscillations. We expect that extracting the signal at much larger extraction radii and extending the signal to null-infinity can alleviate some of these issues. Despite that, one can clearly see, both from Figure \ref{fig:3d_gw_vis} and Figure \ref{fig:bss15c_rpsi4_directions}, that each gravitational-wave burst is sent in very different directions. For other directions, the emission of a burst might be very small, or two peaks that are close to each other might become indistinguishable. For the systems that we present in the following sections, we pick a direction that shows all the existing features of the gravitational-wave signal most clearly. 

\begin{figure*}[htp!]
\includegraphics[width=1.0\columnwidth]{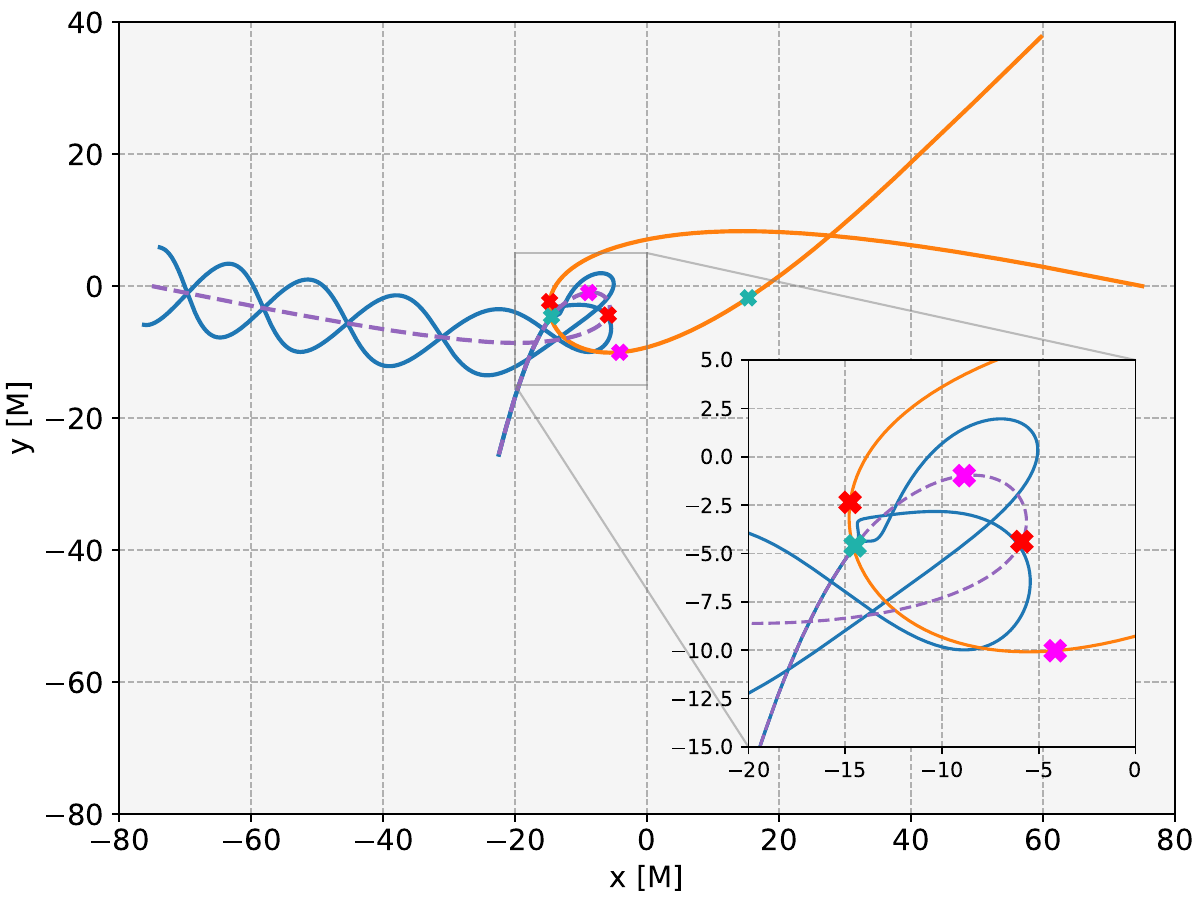}
\includegraphics[width=1.0\columnwidth]{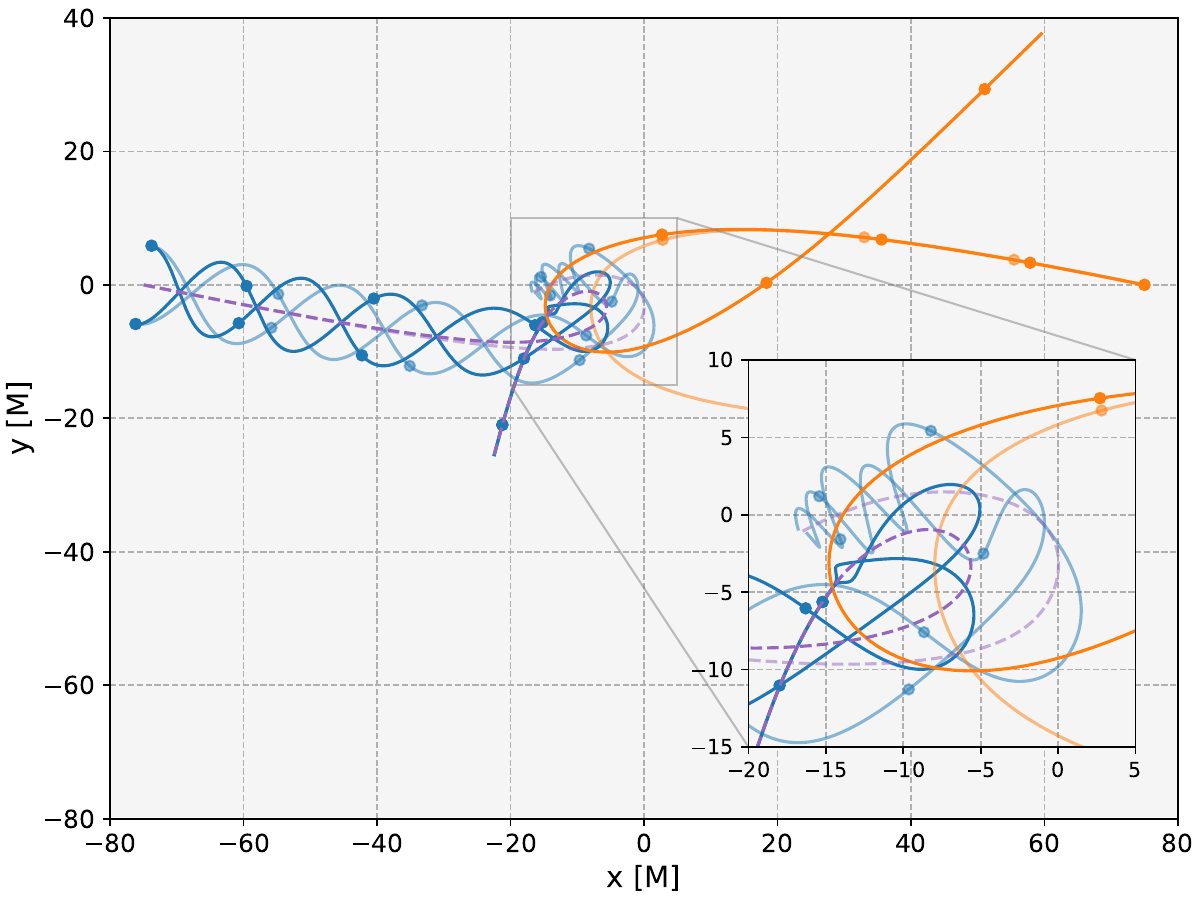}
\caption{Trajectories of the black holes of BSS-03. The single black hole is colored orange, the center of mass of the binary is colored purple and the binary members is colored blue. \textit{Left:} The colored crosses are the positions of the binary's center of mass and the single black hole at the retarded times that correspond to the times marked by the dashed lines in Figure \ref{fig:bss15b_rpsi4m22}. \textit{Right:} Comparison of the fully relativistic simulation (fully opaque) and the 2.5PN simulation (transparent) using the same initial parameters. The positions of the black holes are shown for different time steps separated by 200 M.}
\label{fig:bs_scattering15b_trajectories}
\end{figure*}

\newpage
\subsection{Almost Head-on Collision After a Strong Binary-Single Scattering (BSS-03)}
\label{head_on_collision}

In our third case study, the binary is strongly perturbed by the single black hole such that the black holes in the binary rapidly leave the orbital plane and move towards each other, producing an almost head-on collision. The initial parameters have been obtained by rotating the system of Section \ref{sec:complex_interaction} such that it is set up as described in Section \ref{sec:exp_setup}, with the binary's center of mass and the single black hole initially moving in the $xy$-plane. The vastly different results show that we are in the chaotic regime where small differences in the numerics are enough to change the final outcome. This can be expected when dealing with chaotic systems where the outcome is very sensitive not only to the physical initial conditions but also to the parameters associated with the numerics. 

\begin{figure*}[htp!]
\hspace{5pt}
\includegraphics[height=0.87\columnwidth]{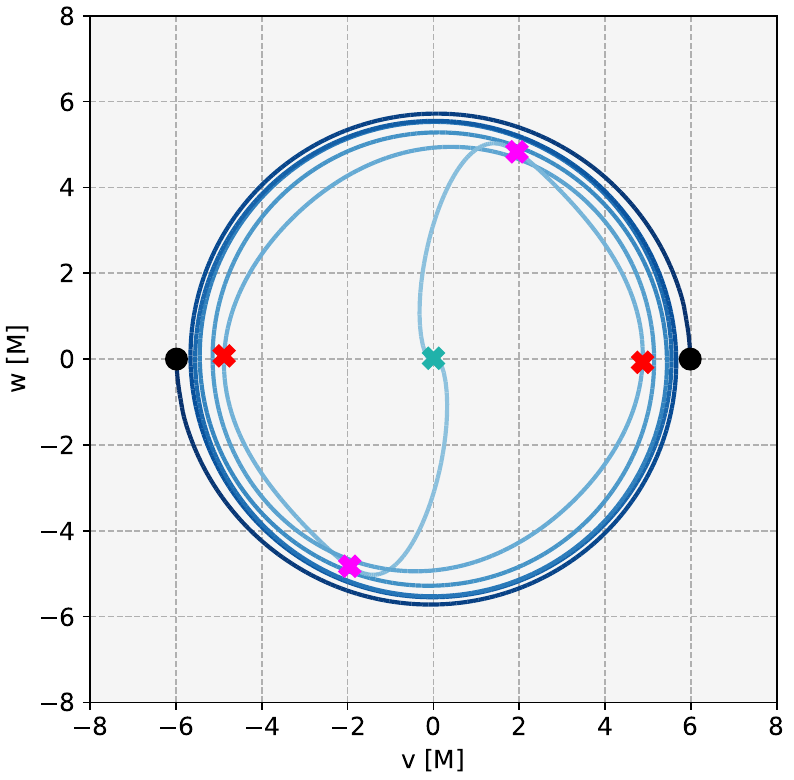}
\hspace{10pt}
\includegraphics[height=0.87\columnwidth]{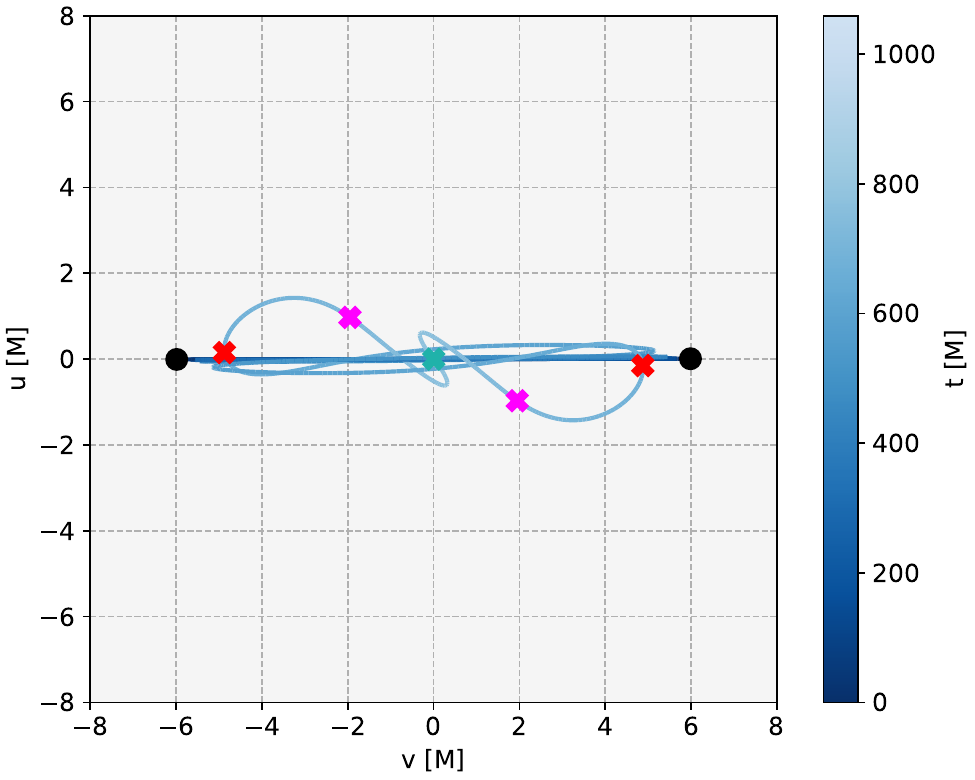}
\caption{Puncture trajectories of the binary black holes of BSS-03 in their center-of-mass frame. The color encodes the positions of the black holes at different times, and the initial positions of the black holes are indicated with black dots. \textit{Left:} Head-on view of the initial orbital plane. \textit{Right:} Edge-on view of the initial orbital plane. The colored crosses are the positions of the black holes at the retarded times that correspond to the times marked by the dashed lines in Figure \ref{fig:bss15b_rpsi4m22}.}
\label{fig:bs_scattering15b_trajectories_com}
\end{figure*}

Figure \ref{fig:bs_scattering15b_trajectories} depicts the black-hole puncture trajectories in the plot on the left, whereas the plot on the right compares the black-hole trajectories of the fully relativistic simulation to the post-Newtonian approximation that uses the same initial parameters. Figure \ref{fig:bs_scattering15b_trajectories_com} shows the puncture trajectories of the binary black holes in their center-of-mass frame. 
\begin{figure*}
    \hspace{-30pt}
    \includegraphics[width=1.0\columnwidth]{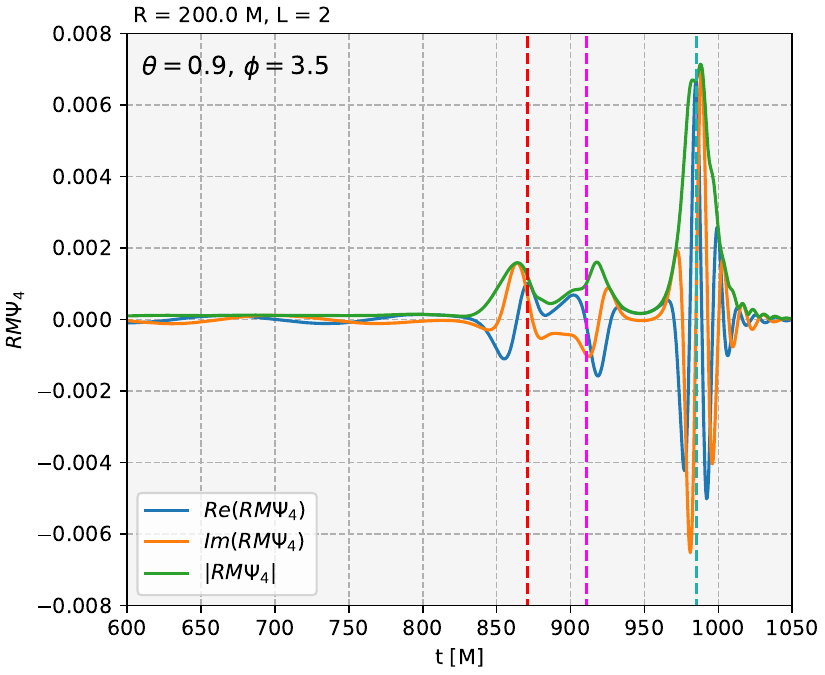}
    \hspace{10pt}
    \includegraphics[width=0.99\columnwidth]{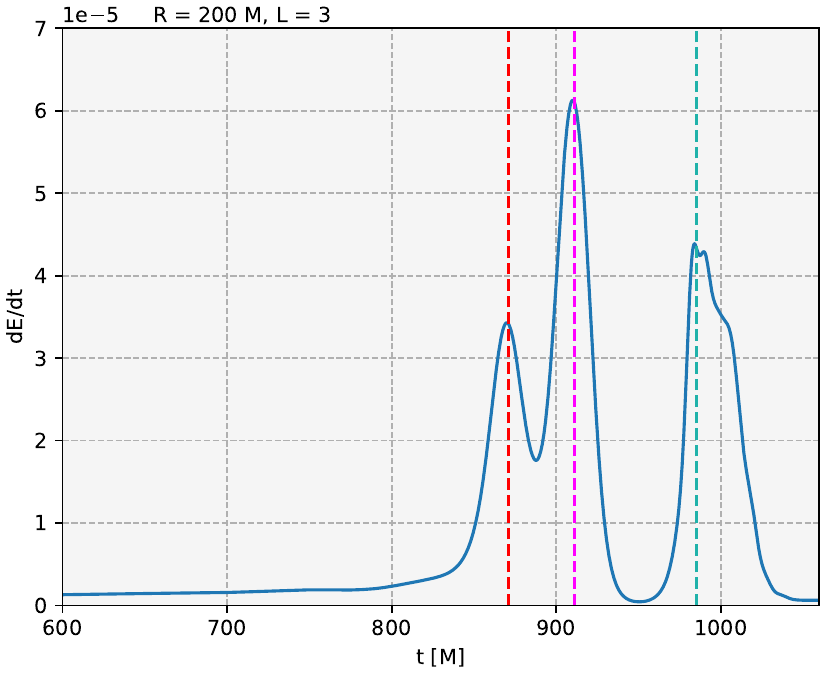}
    \caption{\textit{Left:} The real part (blue), the imaginary part (orange) and the absolute value (green) of $RM\Psi_4$ for the system BSS-03 extracted at the radius $R=200M$ and on the level $L=2$ in the direction $\theta=0.9, \phi=3.5$. $RM\Psi_4$ is reconstructed from all modes from $l=2$ to $l=4$. \textit{Right:} Gravitational-wave luminosity extracted at the same radius and level. The colored dashed lines mark the maxima in gravitational-wave luminosity and correspond to the crosses in Figure \ref{fig:bs_scattering15b_trajectories} and \ref{fig:bs_scattering15b_trajectories_com}.}
    \label{fig:bss15b_rpsi4m22}
\end{figure*}
One can see that the post-Newtonian approximation is much worse than for the system BSS-01 (compare with Figure \ref{fig:nr_25pn_comparison}). In the post-Newtonian simulation, the interaction does not lead to an almost head-on collision but a very eccentric inspiral and the scattering angle is notably smaller, which agrees with the results for the strong-field scattering of individual black holes in Ref. \cite{Damour2014}. The trajectories obtained by numerically integrating the post-Newtonian equations of motion are much less sensitive to the orientation of the system and are therefore the same for the previous system BSS-02 in Section \ref{sec:complex_interaction}.

Figure \ref{fig:bss15b_rpsi4m22} shows $RM\Psi_4$ in a specific direction ($\theta=0.9, \phi=3.5$) as well as the gravitational-wave luminosity, both extracted at the radius $R=200M$ and on the level $L=2$. At the beginning of the simulation, one can see small oscillations in the real and imaginary parts of $RM\Psi_4$ due to the binary inspiral. The insprial is interrupted by three distinct peaks, and the times at which the peak maxima occur are indicated by differently colored dashed lines. The first two gravitational-wave bursts are primarily sent in the positive and negative $z$-directions, whereas the third burst is primarily directed perpendicular to the $z$-axis. Because of this, we choose a direction for $RM\Psi_4$ in which all the peaks that are visible in the gravitational-wave luminosity also clearly appear in the plot of $RM\Psi_4$. For this system, one can also connect the peaks in the gravitational-wave emission to specific events in the binary-single encounter. We again illustrate this connection by adding crosses to the puncture trajectory plots in Figures \ref{fig:bs_scattering15b_trajectories} and \ref{fig:bs_scattering15b_trajectories_com} at the retarded times $t_R = t - R$, which approximately correspond to the moments in time $t$ marked by the dashed lines in Figure \ref{fig:bss15b_rpsi4m22}. One can see that the first peak (colored red) corresponds to the moment when the black holes in the binary get rapidly accelerated out of the orbital plane. This is especially visible in the right plot in Figure \ref{fig:bs_scattering15b_trajectories_com}. 
\begin{figure*}[htp!]
\hspace{5pt}
\includegraphics[height=0.9\columnwidth]{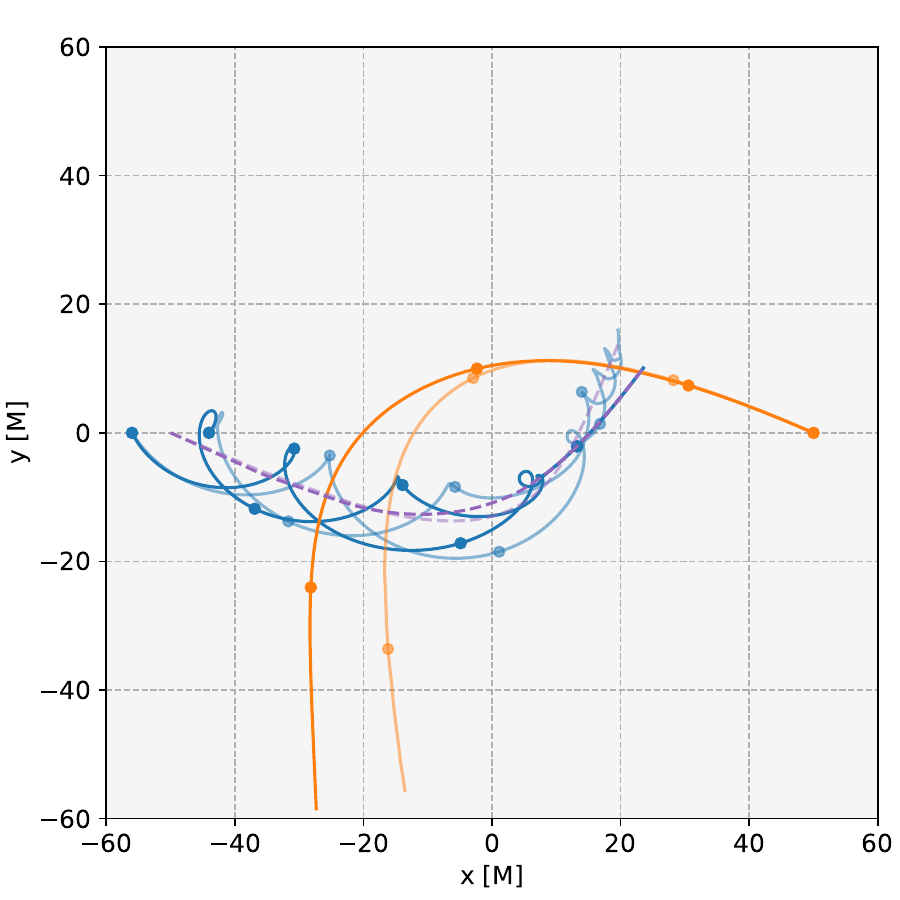}
\hspace{7pt}
\includegraphics[height=0.87\columnwidth]{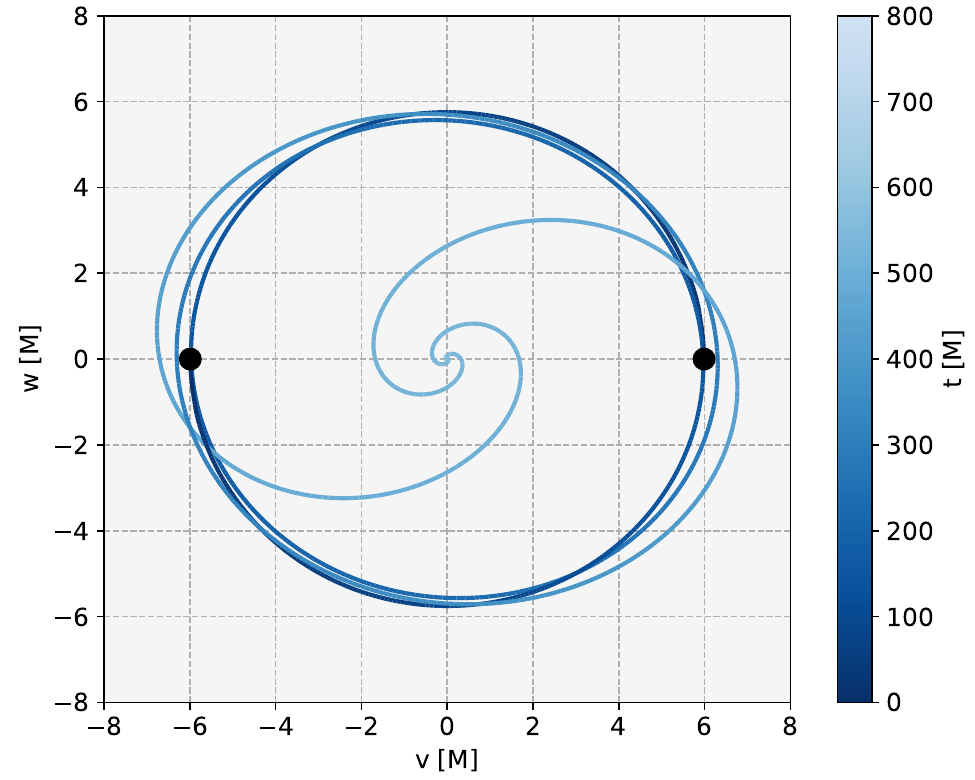}
\caption{Puncture trajectories of the black holes of BSS-04. The trajectory of the single black hole is colored orange, the trajectory of the center of mass of the binary system is colored purple, and the trajectories of the binary members are colored blue. \textit{Left:} Comparison of the fully relativistic simulation (fully opaque) and the post-Newtonian simulation up to order 2.5PN (transparent) using the same initial parameters. The positions of the black holes are shown for different time steps separated by 200 M. \textit{Right:} Head-on view of the orbital plane in the binary's center-of-mass frame. The color encodes the positions of the black holes at different times, and the initial positions of the black holes are indicated with black dots.}
\label{fig:bss18d_trajectories}
\end{figure*}
\begin{figure}[htp!]
\hspace{-20pt}
\includegraphics[width=0.95\columnwidth]{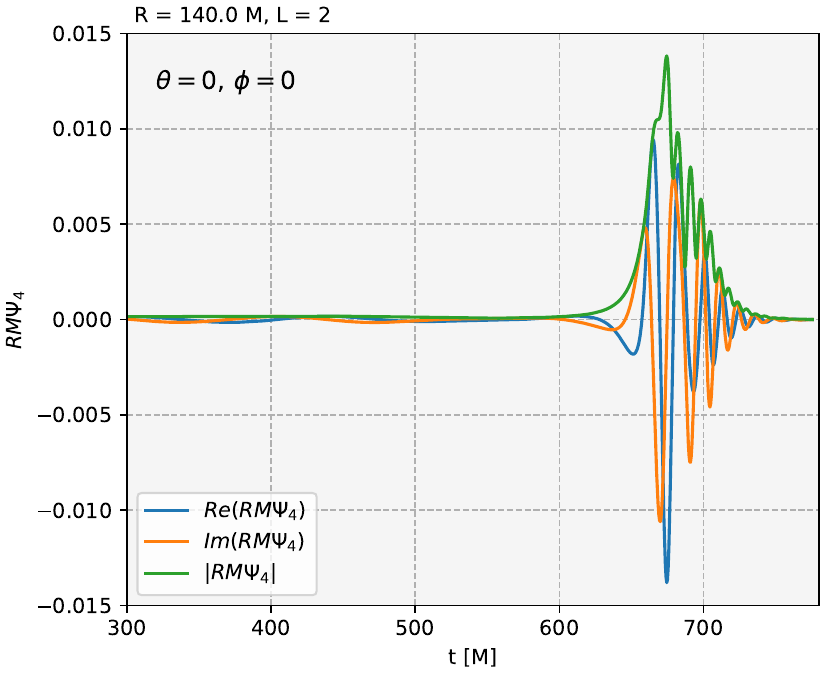}
\caption{The real part (blue), imaginary part (orange), and the absolute value (green) of $RM\Psi_4$ for the system BSS-04, extracted at the radius $R=140M$ and on the level $L=2$ in the direction $\theta=0, \phi=0$ (which corresponds to the $z$-axis).}
\label{fig:bss18d_psi4m22}
\end{figure}
The second peak (colored magenta) corresponds to the moment when the black holes in the binary strongly decelerate, stopping the circular motion entirely and initiating the almost head-on collision, which can be seen most clearly in the left plot in Figure \ref{fig:bs_scattering15b_trajectories_com}. The closest approach of the binary's center of mass and the single black hole is at $t_R = 699M$, which roughly corresponds to the time $t = 899M$ in Figure \ref{fig:bss15b_rpsi4m22}. This time lies in between the first two peaks, so the strong acceleration of the single black hole does not appear to produce a separate, distinct emission peak. The third peak (colored turquoise) corresponds to the merger of the black holes in the binary. In the plot for the gravitational-wave luminosity in Figure \ref{fig:bss15b_rpsi4m22}, this peak has quite an irregular shape, which is likely caused by the associated gravitational-wave burst hitting different points on the extraction sphere at different times. The second peak shows a much higher value of the gravitational-wave luminosity than the actual merger peak, likely because the sudden deceleration of the black holes creates a high second derivative of the mass quadrupole moment, which might be further enhanced by the strong acceleration of the single black hole at that time. We also want to note that the total energy radiated away by the acceleration of the black holes is higher than the energy that is radiated away by the almost head-on merger of the black holes in the binary.

\newpage
\subsection{Highly Eccentric Merger After a Strong Binary-Single Scattering (BSS-04)}
\label{sec:highly_eccentric}

Our fourth system moves in the $xy$-plane with the normal vector of the binary's orbital plane pointing in the $z$-direction. For this system we pick $X=50M$, $b = 40M$ and $|\vec{p}| = 0.057282195M$. The black hole trajectories of the entire system and of the binary in the center-of-mass frame are depicted in Figure \ref{fig:bss18d_trajectories}. Figure \ref{fig:bss18d_psi4m22} shows $RM\Psi_4$ in a specific direction ($\theta=0, \phi=0$, which corresponds to the $z$-axis, i.e., the direction of the normal vector of the binary's orbital plane), extracted at the radius $R=140M$ and on the level $L=2$. One can see that the single black hole induces a significantly accelerated merger with a high eccentricity. In the comparison with the 2.5PN approximation, one can observe that the dynamics look quite similar up to a certain point, although distorted in the direction of motion as in the other examples. The scattering angle appears noticeably larger in the post-Newtonian approximation, which is contrary to what has been observed for the strong-field scattering of individual black holes in \cite{Damour2014}, where the scattering angles of the higher-order post-Newtonian approximations are always smaller than those of the fully relativistic cases.

The gravitational-wave signal in Figure \ref{fig:bss18d_psi4m22} shows the regular motion of the binary at the beginning. This signal gets interrupted and followed by a gravitational-wave burst that is caused by the prompt merger of the black holes in the binary. This example should highlight that a strong binary-single scattering event can also result in a less complex gravitational-wave signal that does not exhibit multiple peaks. The plot of the gravitational-wave luminosity (not shown here) also shows only one regular peak. However, the signal is still very different from a regular merger of a circular binary because of the much sharper transition from the regular binary signal to the gravitational-wave burst of the eccentric merger.

\begin{figure*}[htp!]
\hspace{8pt}
\includegraphics[width=0.79\columnwidth]{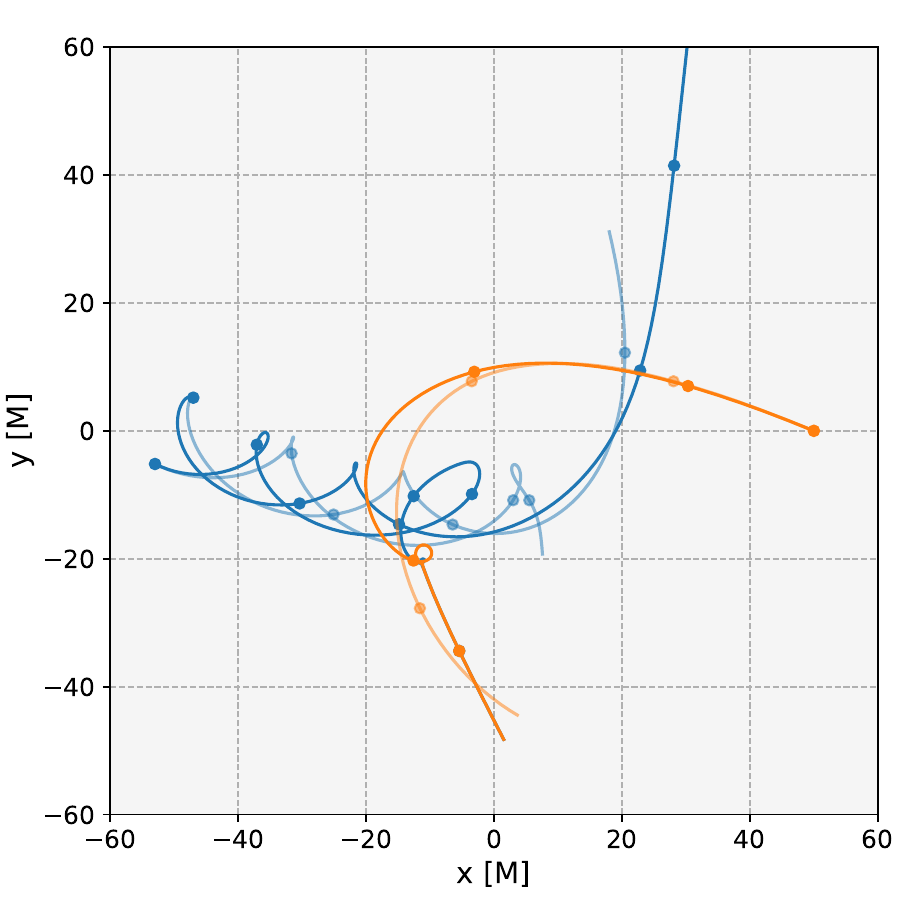}
\hspace{27pt}
\includegraphics[width=0.91\columnwidth]{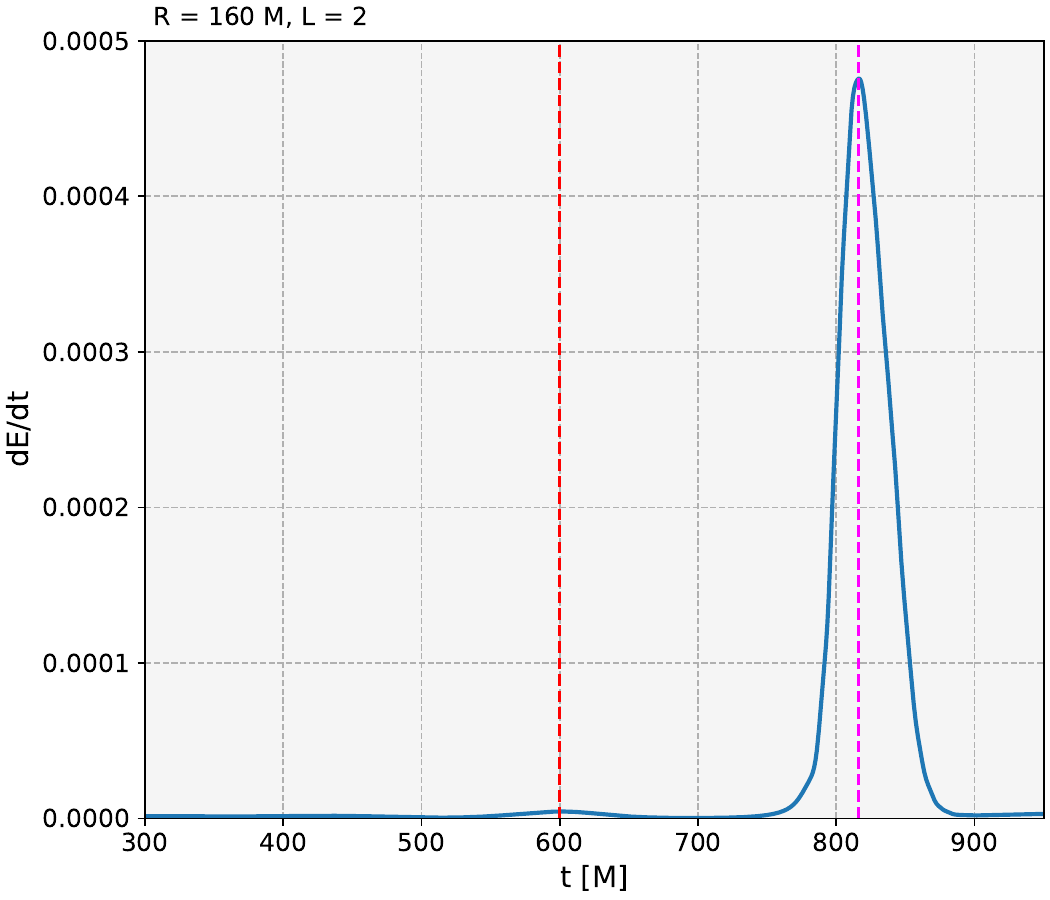}
\hspace{20pt}
\includegraphics[width=0.91\textwidth]{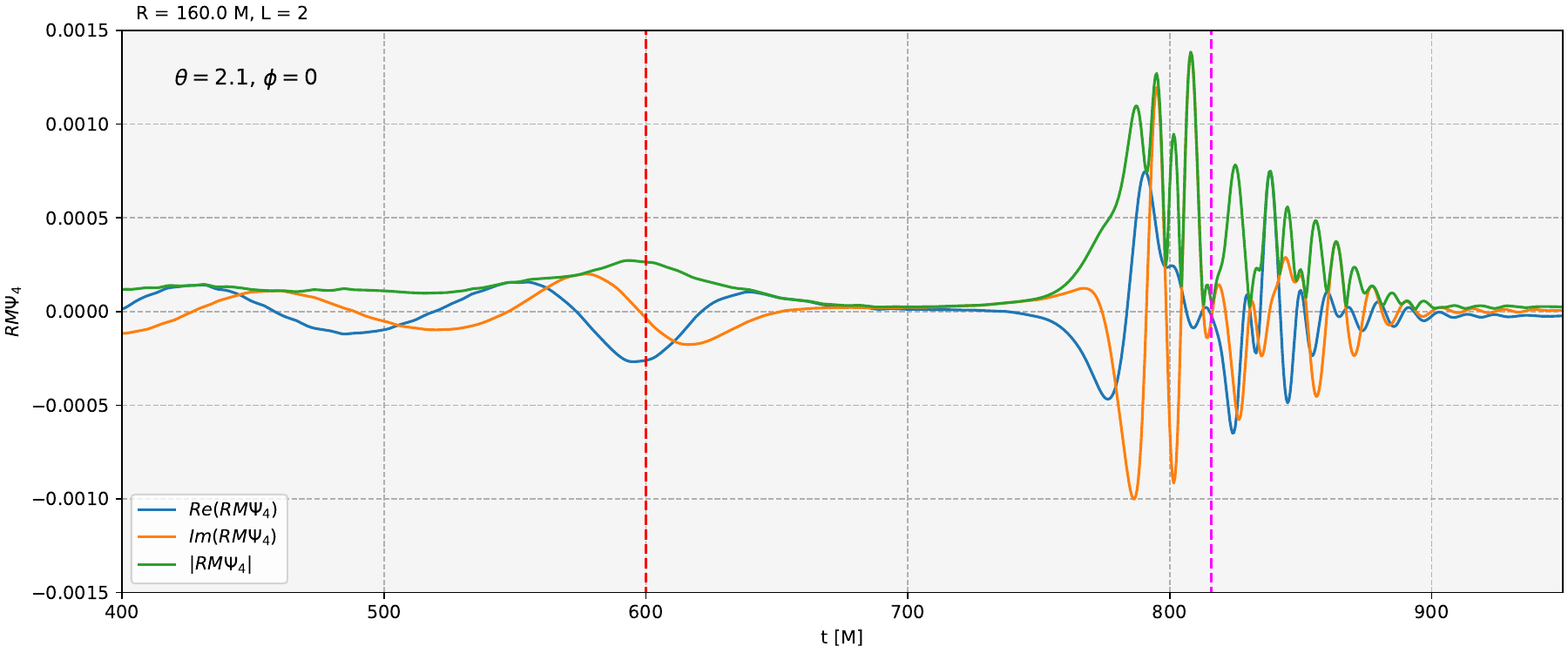}
\caption{\textit{Top left:} Puncture trajectories of the black holes of BSS-05 (colored blue for the black holes in the binary and orange for the single black hole) with a comparison to the 2.5PN approximation. \textit{Top right:} Gravitational-wave luminosity extracted at the radius $R=160M$ and on the level $L=2$. \textit{Bottom:} The real part (blue), the imaginary part (orange), and the absolute value (green) of $RM\Psi_4$, extracted at the same radius and level and in the direction $\theta=2.1, \phi=0$.}
\label{fig:bss24}
\end{figure*}

\subsection{Exchange and Subsequent Merger After a Strong Binary-Single Scattering (BSS-05)}
As a last example of our case studies of black-hole binary-single encounters, we present an exchange immediately followed by a merger while the other binary member gets ejected at a high speed. In this system, all the black holes move in the $xy$-plane with the normal vector of the binary's orbital plane being parallel to the $z$-axis. For this system we choose $X = 50M$, $b = 38M$ and $|\vec{p}| = 0.057282195M$. The detailed list of parameters can be found in Table \ref{tb:parameter_values} in Appendix \ref{sec:extra_material}. The post-Newtonian approximation also produces an exchange, although not immediately followed by a prompt merger. The scattering angle of the remaining objects seems to be larger in the post-Newtonian simulation, which we already observed in the system discussed in Section \ref{sec:highly_eccentric}.
\begin{figure*}[htp!]
\hspace{-20pt}
\includegraphics[height=0.82\columnwidth]{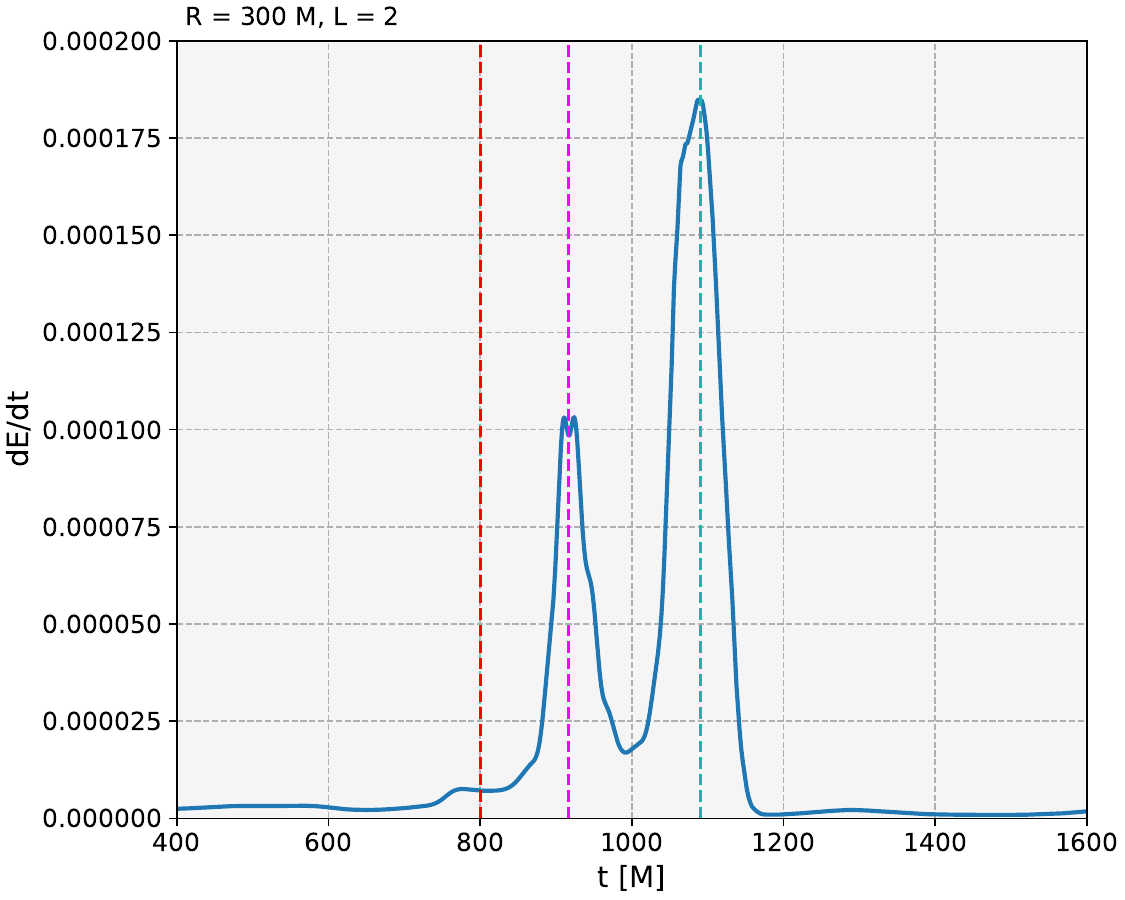}
\includegraphics[height=0.82\columnwidth]{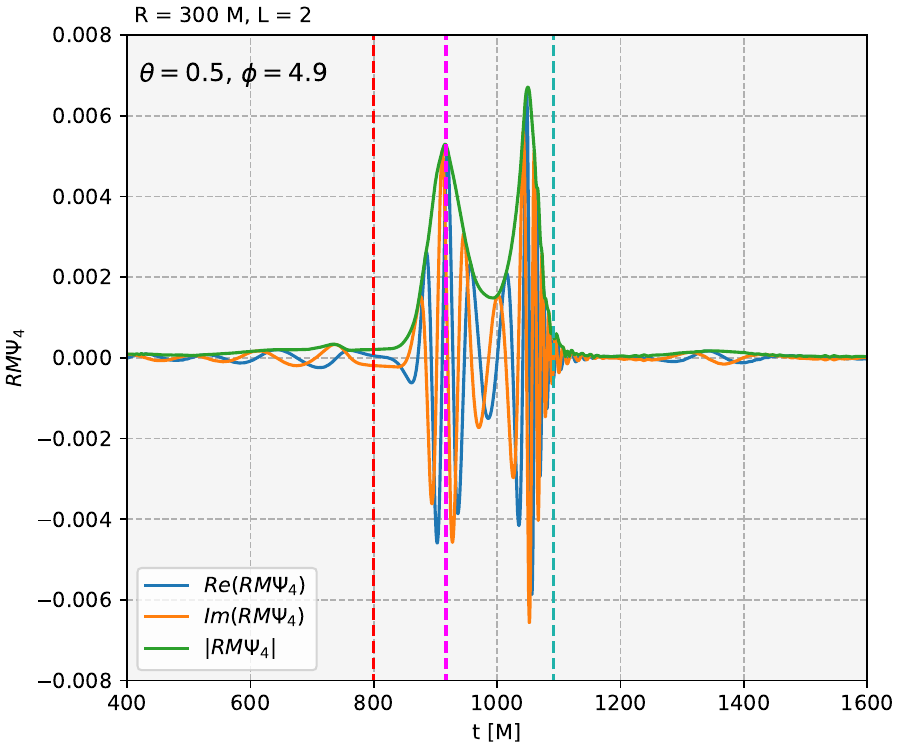}
\caption{\textit{Left:} Gravitational-wave luminosity extracted at the radius $R=300M$ and on the level $L=2$. \textit{Right:} The real part (blue), the imaginary part (orange), and the absolute value (green) of the 22-mode of $RM\Psi_4$, extracted at the same radius and level and in the direction $\theta=0.5, \phi=4.9$.}
\label{fig:bbs03_gw}
\end{figure*}
\begin{figure}
    \centering
    \includegraphics[width=0.95\linewidth]{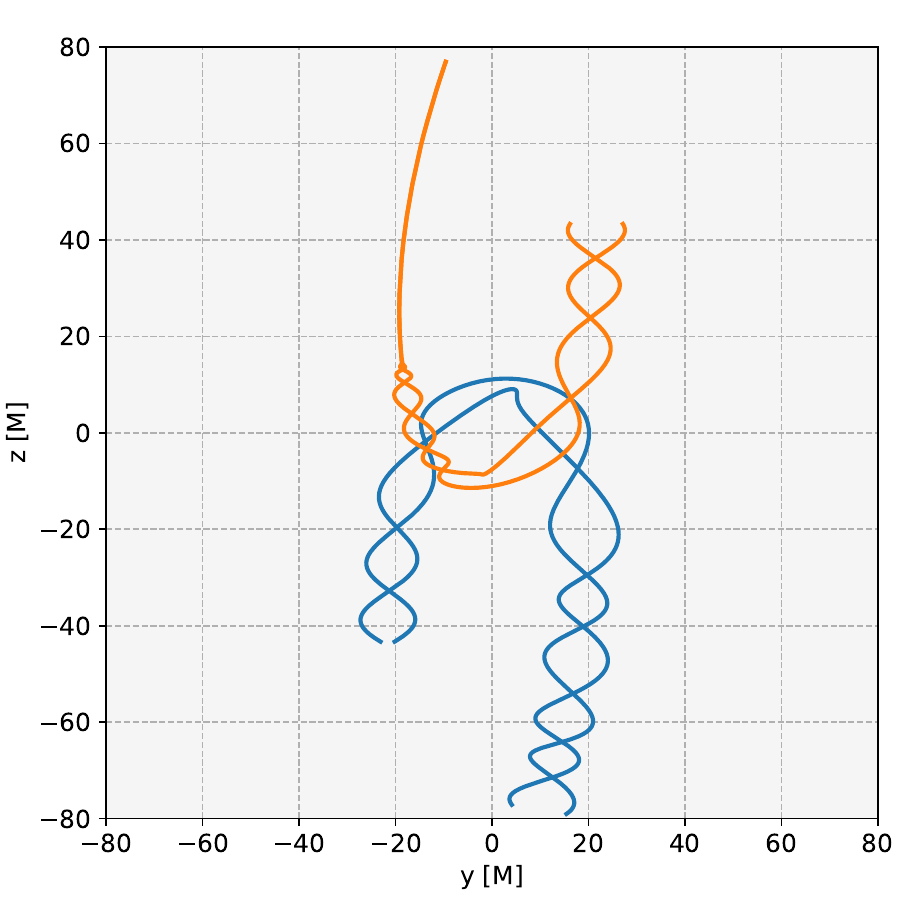}
    \caption{Puncture trajectories of the black holes of BBS-01 with different colors for the different binaries.}
    \label{fig:bbs03_trajectories}
\end{figure}

The plot of $RM\Psi_4$ in Figure \ref{fig:bss24} shows the regular binary signal, interrupted by a small bump and a subsequent decay of the signal, which is then followed by a burst of radiation. This can also be seen in the plot of $dE/dt$ where the bump at $t \approx 600M$ is relatively small. Looking at the system at the retarded time $t_R = t-R$, one can conclude that the small bump arises from the single black hole slowing down one black hole and accelerating the other one, whereas the burst at the end originates from the merger of the single black hole with one of the binary members. It might be the case that the existence of a bump followed by a binary signal with a different frequency, amplitude, and waveform (or a prompt merger as in this case) is a characteristic feature of an exchange since it requires the deceleration of one binary member's circular motion and a strong acceleration of the other binary member, leading to a moderate second derivative of the binary's mass quadrupole moment.

\subsection{Strong Binary-Binary Scattering (BBS-01)}
We now present an example of a strong black-hole binary-binary scattering. For this system we choose $X = 50M$, $b=50M$ and $|\vec{p}| = 0.05M$. Both binaries have the same initial separation, but they start with different phases. We again rotated the system such that the initial orbital plane of the binary is aligned with the $z$-axis and the tetrad basis used to construct $\Psi_4$. 

\begin{figure*}[htp!]
\includegraphics[width=0.96\columnwidth]{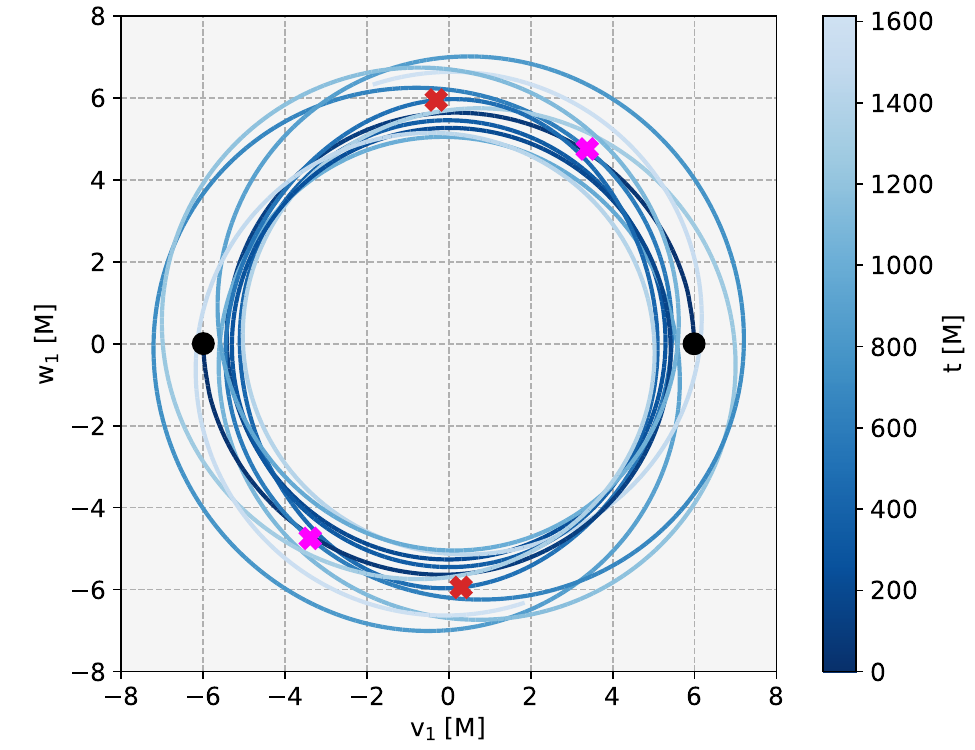}
\hspace{8pt}
\includegraphics[width=0.96\columnwidth]{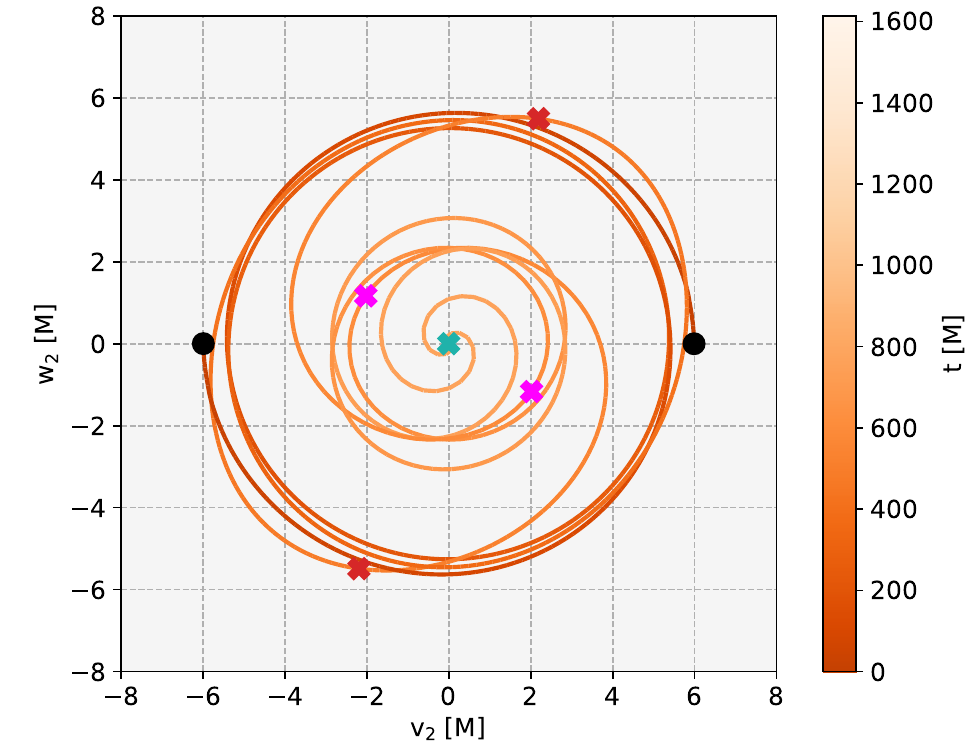}
\caption{Head-on view of the puncture trajectories of the binary black holes of BBS-01 in their center-of-mass frame. The color encodes the positions of the black holes at different times, and the initial positions of the black holes are indicated with black dots. The colored crosses correspond to the times marked by the dashed lines in Figure \ref{fig:bbs03_gw}.}
\label{fig:bbs03_com}
\end{figure*}

Figures \ref{fig:bbs03_trajectories} and \ref{fig:bbs03_com} show that the encounter leads to an accelerated eccentric merger of the binary with orange-colored trajectories, whereas the separation of the black holes in the other binary grows, also leading to an eccentric system. The eccentricity evolution of both systems (not shown here) shows similar characteristics as the one in Figure \ref{fig:bs_scattering02_ecc_evolution} with a high rise at the time around the closest encounter and subsequent oscillations of the eccentricity. 

The plot of $dE/dt$ in Figure \ref{fig:bbs03_gw} shows two distinct and irregularly shaped peaks, as well as a smaller constant emission around the time of closest approach (red dashed line). The first peak (colored magenta) likely corresponds to the strong acceleration of both binaries and/or the acceleration of individual binary members. At the time of the first peak, the black holes in the blue-colored binary get accelerated outwards and reach their widest separation throughout the simulation, while at the same time the black-hole distance in the eccentric orange binary reaches a local minimum. The second peak (colored turquoise) is associated with the merger of the orange-colored binary. One can see that the maximum of the absolute value of $RM\Psi_4$ occurs earlier than in the gravitational-wave luminosity. This is because the merger sends two bursts in opposite directions. One burst reaches the extraction sphere before the time marked by the turquoise line, while the other one hits it afterward. The plot depicting $RM\Psi_4$ also shows the initial signal of both binaries. Due to the different phases, the amplitude gets slightly modulated. The final part of the signal comes from the gravitational-wave emission of the remaining eccentric binary. In the center-of-mass systems of each binary, the black holes get accelerated out of the initial orbital plane during the encounter, and the orbital plane of the binary with blue trajectories gets permanently inclined by around 10 degrees relative to its initial orbital plane.

\section{Summary and Conclusions}
\label{sec:summary}
In this paper, we presented the dynamics and gravitational-wave emission of five fully relativistic black-hole binary-single encounters and one black-hole binary-binary encounter. We compared the dynamics of the binary-single encounters with the predictions from post-Newtonian approximations up to 2.5PN. We started the exploration of the high-dimensional parameter spaces by restricting ourselves to systems with non-spinning, equal-mass black holes and quasi-circular binaries with a fixed initial separation and two possible orientations. The systems show a wide variety of possible outcomes, such as changes in the binary's eccentricity and in the orientation of its orbital plane, delayed, accelerated, or prompt mergers, exchanges, and more complex interactions. 

The PN approximations exhibit similar qualitative dynamics, but the scattering angles can be either smaller or larger compared to the fully relativistic case. For the considered binary's initial separation, the binary phases of the fully relativistic and the post-Newtonian simulation agree well up to several hundred M, but at late times they deviate more strongly, which is mostly due to the inspiral happening faster in the post-Newtonian case. 

We illustrated that in the case of relatively weak binary-single scattering we achieve systematic convergence, while for stronger interactions with smaller impact parameters slight changes in the simulation parameters (including the grid resolution) lead to strong differences in the outcome, which is expected due to the chaotic nature of gravitational $N$-body systems. Weak, mildly relativistic binary-single encounters already show significant deviations from their Newtonian analogs, which mostly have an impact on the scattering angle as well as enhanced motion perpendicular to the plane in which the binary's coordinate center of mass and the single black hole initially move, including weak oscillations with amplitudes on the order of one percent of the binary's separation and a frequency that corresponds to twice the orbital frequency. This perpendicular motion only occurs if the binary's orbital plane is not aligned with the plane in which the binary's center of mass and the single black hole initially move.

The gravitational waves of relativistic binary-single and binary-binary encounters show remarkable characteristics that differ strongly from those of isolated black hole binaries and other sources of gravitational radiation. The signals are mostly characterized by a sequence of sharp peaks with different strengths and a strong directional dependence. These bursts correspond to events of strong accelerations of the binary components, mergers, or the scattering process itself. Additionally, binary-single and binary-binary encounters can lead to weaker bumps as well as changes in the amplitude, frequency, and waveform of the gravitational-wave signal on short time scales due to the changes of the binary's orbital parameters. We have demonstrated that the strong acceleration of the black holes in the binary can emit a stronger burst of gravitational radiation and carry away more energy than the actual merger of the binary, with a higher peak in the rate of change of the system's energy. For a comparison with hyperbolic encounters, dynamical captures and the associated gravitational-wave bursts in systems with two black holes see, e.g., \cite{Damour2014, Gamba2021, Albanesi2024, Fontbute:2024amb}.

The types of systems considered in this study and their corresponding gravitational wave emissions pose significant challenges to current analysis pipelines used to interpret observed signals. On the one hand, existing template-based searches, e.g.\  \cite{Allen:2005fk,Adams:2015ulm,Messick:2016aqy}, which primarily target quasi-circular inspirals, are likely to miss the burst-like gravitational-wave signals associated with more complex dynamical scenarios considered in this work. Hence, detecting such signals would require alternative approaches like coherent burst searches, e.g.~\cite{Klimenko:2015ypf}.

On the other hand, while less `exotic' cases might be captured by existing template banks, accurately interpreting the physical origin of these signals remains a challenge. For instance, none of the current models used in gravitational-wave data analysis account for scenarios where a third body induces increasing orbital eccentricity (cf.~\cite{LIGOScientific:2018mvr,LIGOScientific:2020ibl,KAGRA:2021vkt}), an effect observed in our study. As a result, the extracted source parameters would likely be biased and incorrect. Similarly, consistency checks, those used to test general relativity by comparing the inspiral and ringdown phases, e.g.~\cite{Agathos:2013upa,Brito:2018rfr}, would most likely show inconsistencies indicating a violation of general relativity when, in fact, they are the result of altered merger dynamics introduced by a third object, as demonstrated in several cases we examined.

Altogether, these findings highlight the limitations of existing analysis frameworks and models. Hence, the interpretation of signals from binary-single or binary-binary interactions would, at the current stage, be almost impossible, which raises the hypothetical question of whether, although astrophysically rare, such events happened without being found in the data stream.

In addition to the previously mentioned difficulties, the extraction of the gravitational-wave signal of binary-single and binary-binary encounters is also challenging due to the fact that the binary is not at rest at the center of the extraction sphere. There are works that deal with the signal extraction for binaries whose center of mass is displaced from the origin or moves at a constant velocity \cite{Boyle2016, Chaurasia2018, Woodford2019}, but in our case the binary is accelerating and rotating, which leads to mode mixing and spurious oscillations of $|RM\Psi_4|$. When looking at the gravitational-wave luminosity that takes into account all modes, these oscillations are significantly reduced, but some peaks still show irregular shapes, and it is unclear to what extent this is a physical effect or a result of our method of gravitational-wave signal extraction. 

In general, one needs more sophisticated methods for extracting accurate gravitational waveforms of such complex gravitational $N$-body systems in which the sources constantly move, accelerate, and merge far away from the center of the grid and the extraction sphere. The first step towards improving the gravitational-wave extraction is moving towards larger extraction radii and employing methods that extend the signal to null-infinity. The results presented here should merely show that the signals exhibit interesting and unique characteristics that are not present in other gravitational systems.

The events considered in this study should be quite rare in nature due to the small chance of close encounters at such high speeds right at the moment when the binary is about to merge. However, we believe that systems on larger scales with lower speeds could show similar but much weaker gravitational-wave signals. Especially systems of $N$ supermassive black holes that form through subsequent major galaxy mergers could lead to close binary-single and binary-binary encounters for which a fully relativistic treatment becomes necessary.

In conclusion, a long-term goal of the work initiated here is to establish a sufficiently detailed phenomenology such that encounters of more than two black holes become recognizable in gravitational wave observations, although many challenges remain.

\section*{Acknowledgments}
We thank Anna Neuweiler for very helpful discussions and Gerhard Schäfer for insightful clarifications on the three-body post-Newtonian formalism. 

FMH has been supported by the Deutsche Forschungsgemeinschaft (DFG) under Grant No. 406116891 within the Research Training Group RTG 2522/1. IM and TD acknowledge support from the Deutsche Forschungsgemeinschaft, DFG, project number 504148597 (DI 2553/7). Furthermore, TD acknowledges funding from the EU Horizon under ERC Starting Grant, no.\ SMArt-101076369. Views and opinions expressed are however those of the authors only and do not necessarily reflect those of the European Union or the European Research Council. Neither the European Union nor the granting authority can be held responsible for them. 

The authors gratefully acknowledge the Gauss Centre for Supercomputing e.V. (www.gauss-centre.eu) for providing computing time in project pn36je on the GCS Supercomputer SuperMUC-NG at Leibniz Supercomputing Centre (www.lrz.de). Further computations were performed on the Ara cluster at the Friedrich-Schiller University Jena, funded by the Deutsche Forschungsgemeinschaft (DFG, German  Research Foundation) - 273418955 and 359757177.

\section*{Data Availability}
Animated videos of some of the plots presented in this paper can be found at \url{https://doi.org/10.5281/zenodo.15342415} \cite{Heinze2025}. They give valuable insights that are not obvious from the plots. The animated version of Figure \ref{fig:3d_gw_vis} was created with rose \cite{markin_2025_14627566}. The code for integrating the three-body post-Newtonian equations of motion up to 2.5PN is available at \url{https://github.com/fmheinze/pn-nbody} \cite{pn-nbody}. All raw data corresponding to the findings in this manuscript are not publicly available upon publication because it is not technically feasible and/or the cost of preparing, depositing, and hosting the data would be prohibitive within the terms of this research project. The data are available from the authors upon reasonable request.

\appendix

\begin{table*}[htp!]
\begin{tabular}{l|l|l|l|l|l|l|l}
\hline
\multicolumn{1}{|l|}{Name} & BSS-01a & BSS-01b & BSS-02 & BSS-03 & BSS-04 & BSS-05 & \multicolumn{1}{l|}{BBS-01} \\ \hline \hline

\multicolumn{1}{|l|}{\textbf{$b/M$}} & 50.0 & 50.0 & 30.0 & 30.0 & 40.0 & 38.0 & \multicolumn{1}{l|}{50.0} \\ \hline
\multicolumn{1}{|l|}{\textbf{$X/M$}} & 100.0 & 50.0 & 75.0 & 75.0 & 50.0 & 50.0 & \multicolumn{1}{l|}{50.0}  \\ \hline
\multicolumn{1}{|l|}{\textbf{$|\vec{p}|/M$}} & 0.057282195 & 0.057282195 & 0.05 & 0.05 & 0.057282195 & 0.057282195 & \multicolumn{1}{l|}{0.05}  \\ \hline \hline

\multicolumn{1}{|l|}{\textbf{$m_1/M$}} & 0.487398 & 0.487398 & 0.487669 & 0.487669 & 0.487669 & 0.486589 & \multicolumn{1}{l|}{0.48775}  \\ \hline
\multicolumn{1}{|l|}{\textbf{$m_2/M$}} & 0.487398 & 0.487398 & 0.487669 & 0.487669 & 0.487669 & 0.486589 & \multicolumn{1}{l|}{0.48775}  \\ \hline
\multicolumn{1}{|l|}{\textbf{$m_3/M$}} & 0.487398 & 0.487398 & 0.487669 & 0.487669 & 0.487669 & 0.486589 & \multicolumn{1}{l|}{0.48775}  \\ \hline
\multicolumn{1}{|l|}{\textbf{$m_4/M$}} & - & - & - & - & - & - & \multicolumn{1}{l|}{0.48775}  \\ \hline \hline

\multicolumn{1}{|l|}{\textbf{$x_1/M$}} & -98.503525 & -47.00705 & -1.80282 & -73.80282 & -44.0141 & -47.007050 & \multicolumn{1}{l|}{-18.30384516} \\ \hline
\multicolumn{1}{|l|}{\textbf{$y_1/M$}} & 5.79582275 & 5.18394146 & -8.8319782 & 5.86496026 & 0.0 & 5.183941 & \multicolumn{1}{l|}{-23.11568805}  \\ \hline
\multicolumn{1}{|l|}{\textbf{$z_1/M$}} & 0.0 & 0.0 & -73.48469228 & 0.0 & 0.0 & 0.0 & \multicolumn{1}{l|}{-43.30127019}  \\ \hline
\multicolumn{1}{|l|}{\textbf{$x_2/M$}} & -101.496475 & -52.99295 & -4.19718 & -76.19718 & -55.9859 & -52.992950 & \multicolumn{1}{l|}{-6.69615484} \\ \hline
\multicolumn{1}{|l|}{\textbf{$y_2/M$}} & -5.79582275 & -5.18394146 & -20.56189872 & -5.86496026 & 0.0 & -5.183941 & \multicolumn{1}{l|}{-20.18558214}  \\ \hline
\multicolumn{1}{|l|}{\textbf{$z_2/M$}} & 0.0 & 0.0 & -73.48469228 & 0.0 & 0.0 & 0.0 & \multicolumn{1}{l|}{-43.30127019}  \\ \hline
\multicolumn{1}{|l|}{\textbf{$x_3/M$}} & 100.0 & 50.0 & 3.0 & 75.0 & 50.0 & 50.0 & \multicolumn{1}{l|}{10.11704545} \\ \hline
\multicolumn{1}{|l|}{\textbf{$y_3/M$}} & 0.0 & 0.0 & 14.69693846 & 0.0 & 0.0 & 0.0 & \multicolumn{1}{l|}{16.15950348}  \\ \hline
\multicolumn{1}{|l|}{\textbf{$z_3/M$}} & 0.0 & 0.0 & 73.48469228 & 0.0 & 0.0 & 0.0 & \multicolumn{1}{l|}{43.30127019}  \\ \hline
\multicolumn{1}{|l|}{\textbf{$x_4/M$}} & - & - & - & - & - & - & \multicolumn{1}{l|}{14.88295455} \\ \hline
\multicolumn{1}{|l|}{\textbf{$y_4/M$}} & - & - & - & - & - & - & \multicolumn{1}{l|}{27.14176671}  \\ \hline
\multicolumn{1}{|l|}{\textbf{$z_4/M$}} & - & - & - & - & - & - & \multicolumn{1}{l|}{43.30127019}  \\ \hline \hline

\multicolumn{1}{|l|}{\textbf{$p_{x,1}/M$}} & 0.05536625 & 0.04941384 & -0.08352486 & 0.04891219 & 0.052112 & -0.020966 & \multicolumn{1}{l|}{0.02122112} \\ \hline
\multicolumn{1}{|l|}{\textbf{$p_{y,1}/M$}} & -0.01469623 & -0.02897712 & 0.01665344 & -0.01038016 & 0.06225512 & 0.020481 & \multicolumn{1}{l|}{-0.08248274}  \\ \hline
\multicolumn{1}{|l|}{\textbf{$p_{z,1}/M$}} & 0.085168 & 0.085168 & 0.05 & 0.085168 & 0.0 & 0.0 & \multicolumn{1}{l|}{0.05}  \\ \hline
\multicolumn{1}{|l|}{\textbf{$p_{x,2}/M$}} & 0.05556025 & 0.04980184 & 0.08352486 & 0.04906739 & 0.052888 & 0.126937 & \multicolumn{1}{l|}{-0.02122112} \\ \hline
\multicolumn{1}{|l|}{\textbf{$p_{y,2}/M$}} & -0.01394487 & -0.02830508 & -0.01665344 & -0.00961984 & -0.10808088 & -0.064015 & \multicolumn{1}{l|}{0.08248274}  \\ \hline
\multicolumn{1}{|l|}{\textbf{$p_{z,2}/M$}} & -0.085168 & -0.085168 & 0.05 & -0.085168 & 0.0 & 0.0 & \multicolumn{1}{l|}{0.05}  \\ \hline
\multicolumn{1}{|l|}{\textbf{$p_{x,3}/M$}} & -0.05546325 & -0.04960784 & 0.0 & -0.04898979 & -0.0525 & -0.052985 & \multicolumn{1}{l|}{0.07828285} \\ \hline
\multicolumn{1}{|l|}{\textbf{$p_{y,3}/M$}} & 0.01432055 & 0.0286411 & 0.0 & 0.01 & 0.02291288 & 0.021767 & \multicolumn{1}{l|}{-0.03354899}  \\ \hline
\multicolumn{1}{|l|}{\textbf{$p_{z,3}/M$}} & 0.0 & 0.0 & -0.05 & 0.0 & 0.0 & 0.0 & \multicolumn{1}{l|}{-0.05}  \\ \hline
\multicolumn{1}{|l|}{\textbf{$p_{x,4}/M$}} & - & - & - & - & - & - & \multicolumn{1}{l|}{-0.07828285} \\ \hline
\multicolumn{1}{|l|}{\textbf{$p_{y,4}/M$}} & - & - & - & - & - & - & \multicolumn{1}{l|}{0.03354899}  \\ \hline
\multicolumn{1}{|l|}{\textbf{$p_{z,4}/M$}} & - & - & - & - & - & - & \multicolumn{1}{l|}{-0.05}  \\ \hline

\end{tabular}
\caption{Parameter values for all the binary-single and binary-binary encounters presented in the main text.}
\label{tb:parameter_values}
\end{table*}

\begin{figure*}[htp!]
    \centering
    \hspace{-30pt}
    \includegraphics[width=0.9\linewidth]{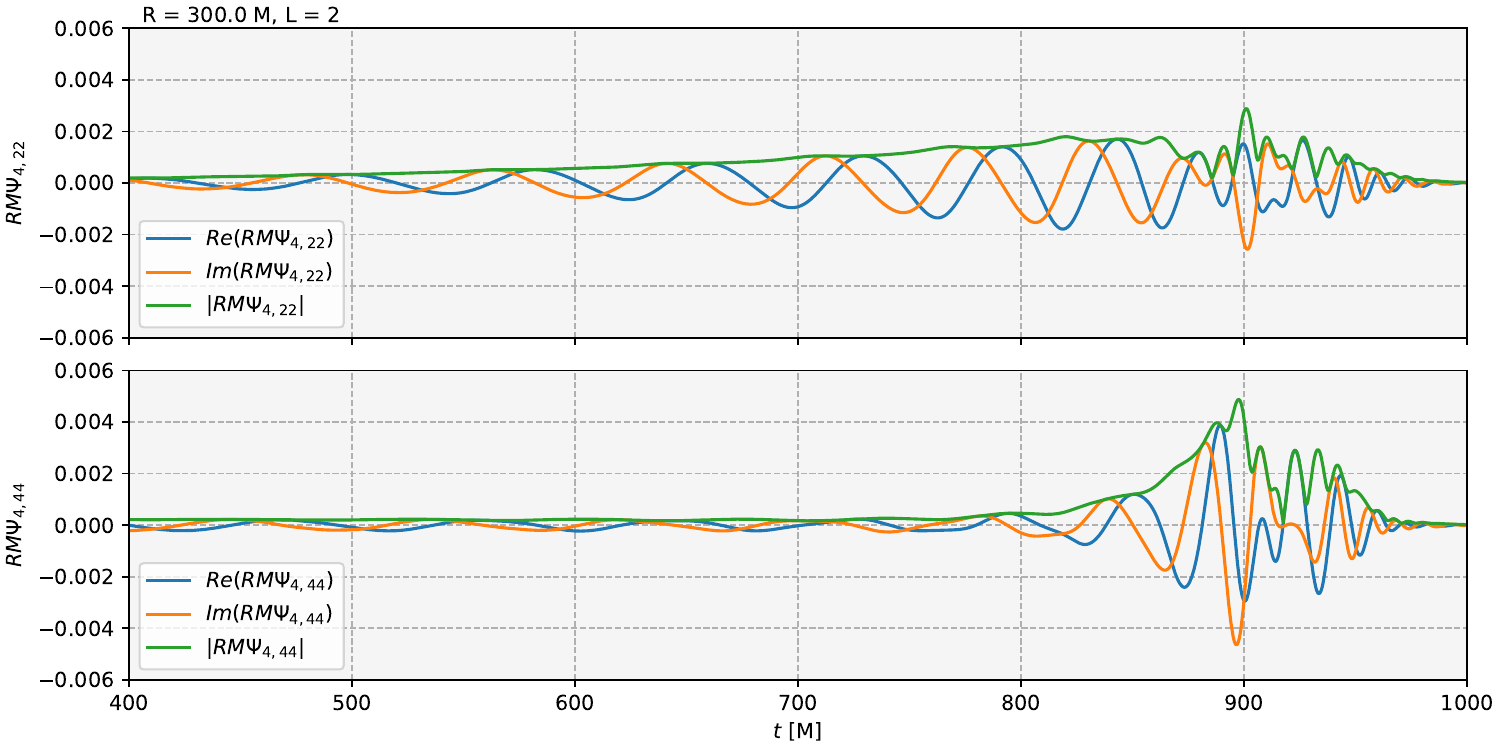}
    \caption{The real part (blue), the imaginary part (orange), and the absolute value (green) of the 22-mode (top panel) and the 44-mode (bottom panel) of $RM\Psi_4$ for a quasi-circular black-hole binary moving with a momentum $|\vec{p}|=0.05M$ in the positive $x$-direction, away from the coordinate origin. The signal is extracted at the radius $R = 300M$ and on the level $L = 3$.}
    \label{fig:gw_bbh_moving}
\end{figure*}

\begin{figure*}
\includegraphics[width=1.0\columnwidth]{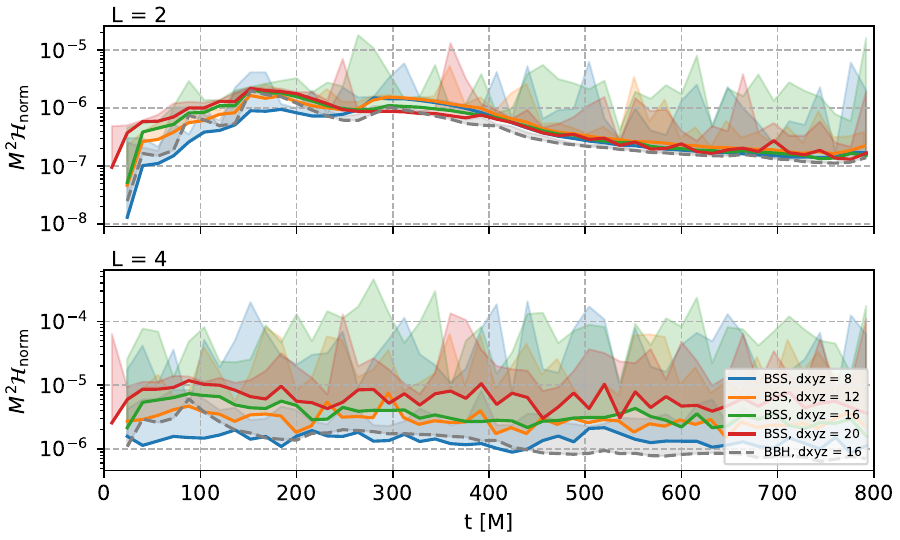}
\hspace{8pt}
\includegraphics[width=1.0\columnwidth]{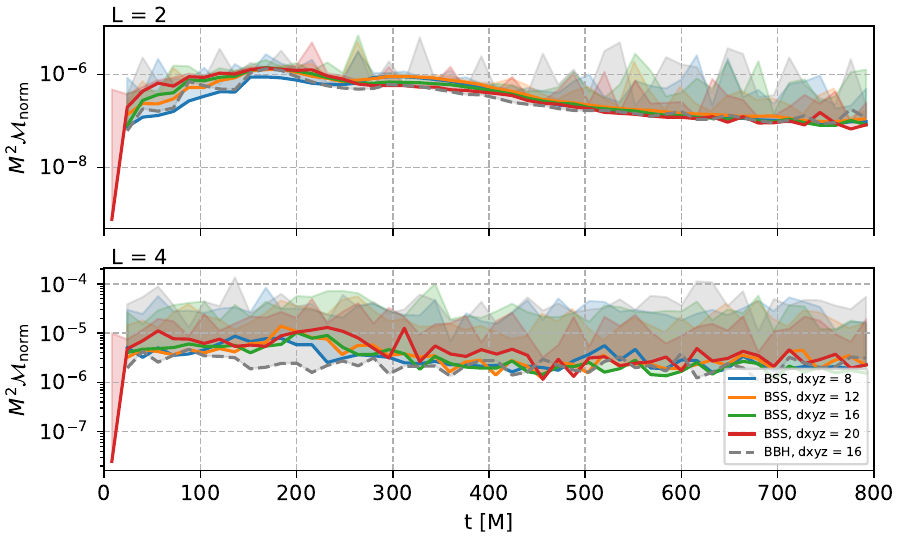}\vspace{10pt}
\includegraphics[width=1.0\columnwidth]{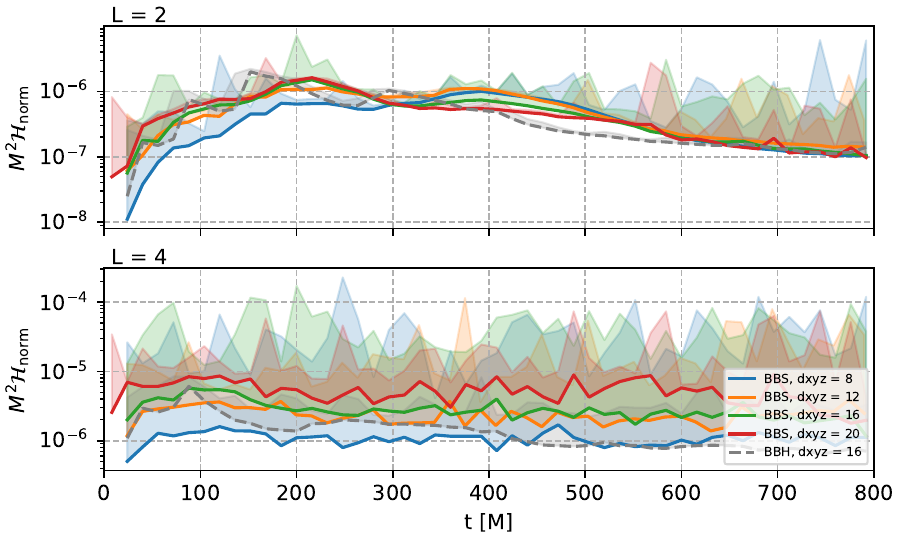}
\hspace{8pt}
\includegraphics[width=1.0\columnwidth]{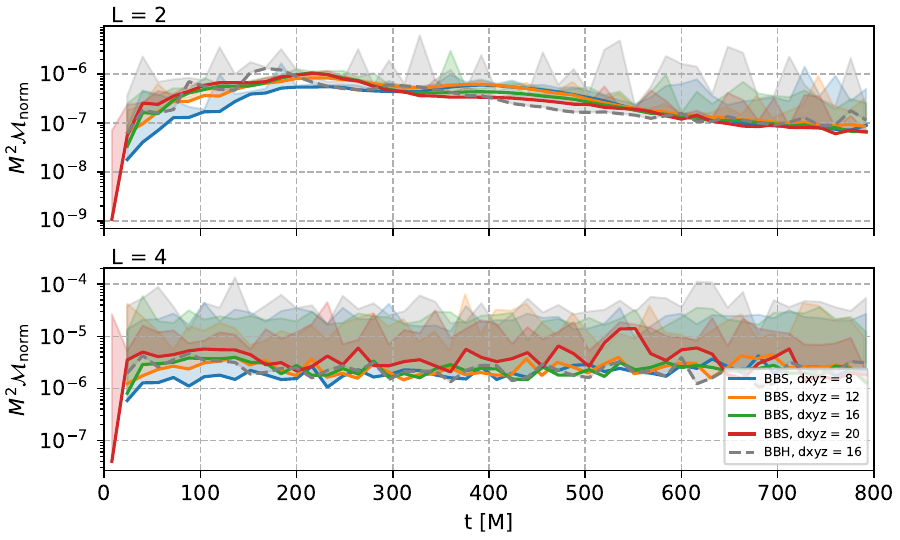}
\caption{The $l2$ norm of the Hamiltonian constraint (left column) and momentum constraint (right column) of the system BSS-01b (top panels) and BBS-01 (bottom panels) for different resolutions and on two different non-moving levels. For comparison, the gray line indicates the values of a black-hole binary which is fixed at the center of the computational domain. The data is binned in time and the transparent filled regions indicate the total value range for each bin.}
\label{fig:constraints}
\end{figure*}

\begin{figure}
    \centering
    \includegraphics[width=1.0\linewidth]{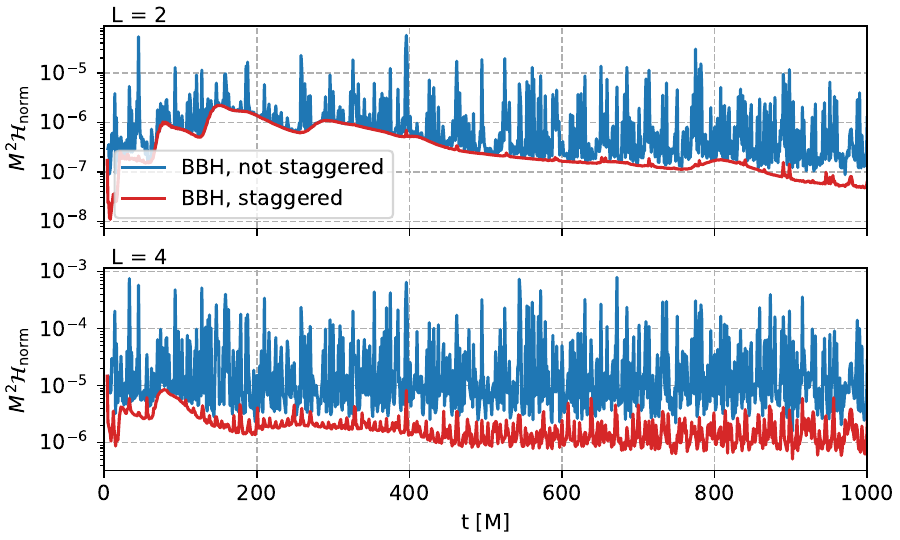}
    \caption{Comparison of the Hamiltonian constraint for two binary systems, where one is staggered by half a grid spacing and the other is not staggered.}
    \label{fig:ham_constraint_bbh}
\end{figure}

\section{Extra Material}
\label{sec:extra_material}

Table \ref{tb:parameter_values} provides the full list of parameters used for all the systems in the main text. In Figure \ref{fig:gw_bbh_moving} we show the 22-mode and the 44-mode of $RM\Psi_4$ for a quasi-circular binary that is moving uniformly with respect to the origin of the gravitational-wave extraction sphere as a simple example of mode mixing and the associated spurious oscillations that are discussed in the main text. 

\begin{figure*}[htp!]
    \centering
    \includegraphics[width=1\textwidth]{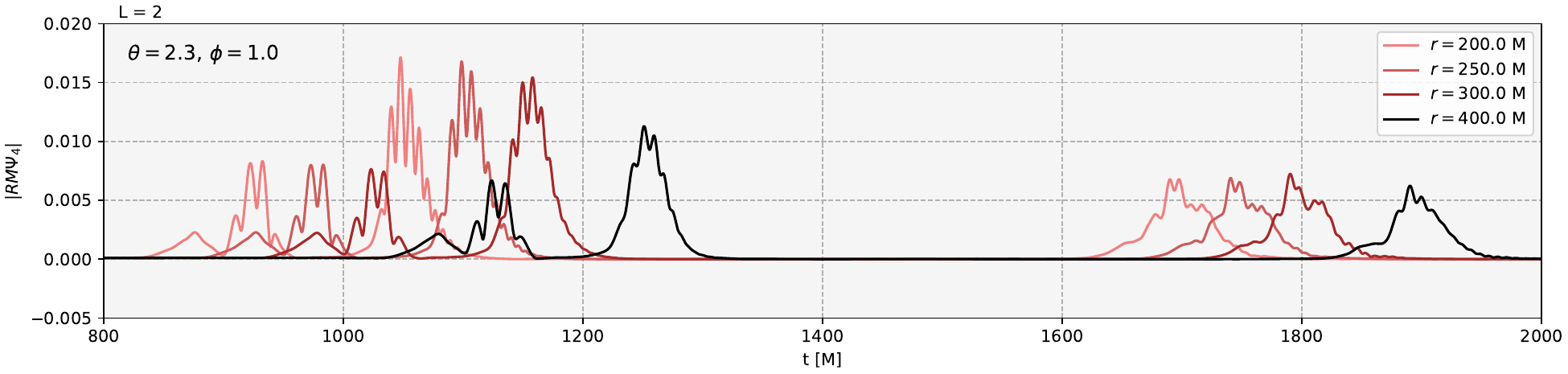}
    \caption{The absolute value of $RM\Psi_4$ for the system BSS-02 extracted at different radii (indicated with different colors) on the level $L=2$ and in the direction $\theta=2.3, \phi=1.0$. $RM\Psi_4$ is reconstructed from all modes from $l=2$ to $l=4$.}
    \label{fig:gw_radii}
\end{figure*}
\begin{figure}[htp!]
    \centering
    \includegraphics[width=1.0\columnwidth]{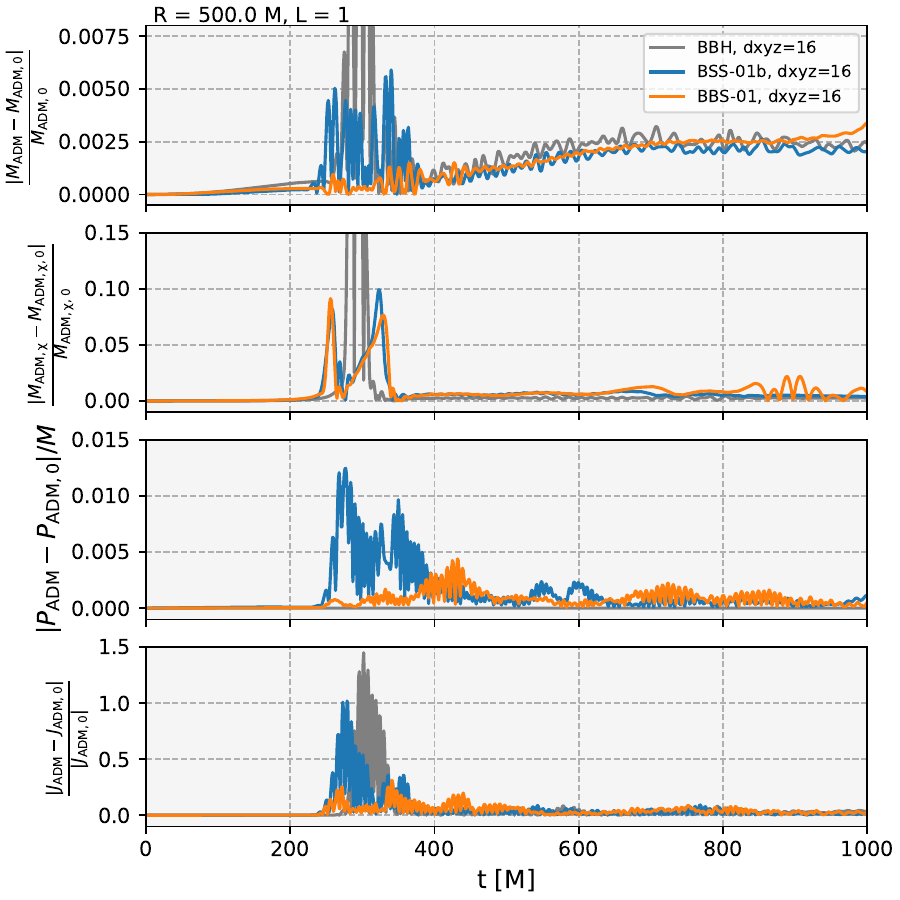}
    \caption{Time evolution of the ADM integrals for the mass (first two panels), the $z$-component of the linear momentum (third panel) and the angular momentum (last panel) for the system BSS-01b (solid orange lines) and black-hole binary that is fixed at the center of the grid (dashed gray lines).}
    \label{fig:adm_conservation}
\end{figure}
\section{Numerical Accuracy}
\label{sec:numerical_accuracy}
In order to test the numerical accuracy of our simulations, we computed the $l2$ norm of the Hamiltonian constraint and the momentum constraint violation for a representative binary-single and binary-binary system. In Figure \ref{fig:constraints} one can see these quantities plotted and compared with those of a black-hole binary that is fixed at the center of the computational domain (using initial parameters from \cite{Tichy2011}) and which uses the same numerical setup. For a cleaner visualization, the data is binned in time, and the transparent filled regions indicate the total value range for each bin. The unbinned data look like a baseline (corresponding to the solid lines at the bottom of each shaded region) with several strong spikes that likely occur when a puncture comes close to a grid point. We come to this conclusion because the number of spikes increases with level, resolution, and the number of black holes. Note that we show the norm of the constraints without further normalization, e.g.\ by second derivatives of the metric, so constraint violations near the punctures are clearly visible. 
To examine the effect of the distance between punctures and grid points, we show two simulations for the simple binary, where in one case the orbital plane is staggered between the grid points (the default in \textsc{BAM}, as in Figure \ref{fig:constraints}), and in the other case without staggering.
The spikes strongly increase when the punctures lie in a plane of grid points and therefore come closer to grid points during the run (see Figure \ref{fig:ham_constraint_bbh}). Apart from that, one can see that the constraint violations have a constant to decreasing trend over time. For the Hamiltonian constraint, one can see an improvement with higher resolution, especially when going to $\mathrm{dxyz}=8$. For the momentum constraint, one can only see an improvement with higher resolution for early times. In all cases, the constraint violations for $\mathrm{dxyz}=16$ are comparable or larger by a factor of a few compared to those of the simple binary system. It would be interesting to consider constraint damping as in \cite{Etienne2024}.

In Figure \ref{fig:gw_radii} we investigate the consistency of the extracted gravitational-wave signals by plotting $|RM\Psi_4|$ for different extraction radii. The relative positions and the approximate shapes of the peaks do not vary much. One can, however, observe that the height of distinct peaks changes differently with radius. This leads to the relative heights of the peaks changing with the extraction radius. Additionally, the spurious oscillations vary slightly for different extraction spheres and they can decrease in amplitude (see e.g.\ the highest peak in Figure \ref{fig:gw_radii}). All of this does not significantly change when considering more modes up to $l=6$. Besides the $lm$-modes of $\Psi_4$ one also has to consider that under certain conditions, there can be a mixing of the Penrose-Newman scalars $\Psi_n$ themselves. The different falloff behaviors of the Penrose-Newman scalars could potentially explain some of the different falloff behaviors observed in our gravitational wave signals. Since the phenomenology and the qualitative aspects of the signals do not change and are consistent across different extraction spheres, and since they all correspond to strong gravitational interactions in the corresponding systems, we consider our statements in the main text to be unaffected by this.

We also tested the conservation of mass, linear momentum, and angular momentum, which is analyzed by monitoring the corresponding ADM surface integrals. This is illustrated in Figure \ref{fig:adm_conservation} for the binary-single system BSS-01b and the binary-binary system BBS-01, where the relative deviations form the initial values are plotted. These are compared to those of a simple black-hole binary. Initially, one can see strong variations due to junk radiation from imperfect initial data. These variations are the smallest for the binary-binary system. After that, the qualitative behavior of the quantities of the considered systems are very similar. In general, the binary system has the best conservation properties throughout the simulation. The binary-binary system shows slightly worse conservation properties than the binary-single system at late times, especially for the conservation of mass-energy. This is likely due to the more dynamical nature of systems with more black holes, which are not restricted to motion within a plane.

\clearpage
\bibliography{apssamp}

\end{document}